\documentclass[11pt,a4paper]{article}
\usepackage{jheppub}
\usepackage{amsmath}
\usepackage{amssymb} 
\usepackage{booktabs} 
\numberwithin{equation}{section}

\usepackage{url}
\usepackage{hyperref}
\usepackage{graphicx}
\usepackage{xspace}
\usepackage[T1]{fontenc}

\newcommand{\cM}{{\cal M}}

\newcommand{\cF}{{\cal F}}

\newcommand{\jet}{\text{jet}}
\newcommand{\Higgs}{\text{Higgs}}

\newcommand{\MSbar}{\overline{\mbox{\scriptsize MS}}}

\newcommand{\ptjv}{p_{\rm t,veto}}

\newcommand{\ptm}{p_{\rm t,j1}}
\newcommand{\ptM}{p_{\rm t, veto}^{\rm Max}}
\newcommand{\eff}{\epsilon}
\newcommand{\pt}{p_{\rm t}}
\newcommand{\MCatNLO}{\textsc{mc@nlo}\xspace}
\newcommand{\POWHEG}{\textsc{powheg}\xspace}
\newcommand{\Pythia}{\textsc{pythia}\xspace}
\newcommand{\PWGPYT}{\textsc{powheg+pythia}\xspace}
\newcommand{\HqT}{\textsc{h}q\textsc{t}\xspace}
\newcommand{\CAESAR}{\textsc{caesar}\xspace}
\newcommand{\FEHIP}{\textsc{fehip}\xspace}
\newcommand{\FEWZ}{\textsc{fewz}\xspace}
\newcommand{\HNNLO}{\textsc{hnnlo}\xspace}
\newcommand{\DYNNLO}{\textsc{dynnlo}\xspace}

\def\TR{T_R}
\def\CA{C_A}

\def\nf{n_{\!f}}

\newcommand{\ee}{e^+e^-}

\newcommand{\order}[1]{{\cal O}\left(#1\right)}

\newcommand{\as}{\alpha_s}
\newcommand{\asCMW}{\alpha_s^{\rm CMW}}

\newcommand{\GeV}{\;\mathrm{GeV}}
\newcommand{\TeV}{\;\mathrm{TeV}}

\newcommand{\tw}{\textwidth}

\def\cO#1{{\cal{O}}\left(#1\right)}

\newcommand{\Ltilde}{\widetilde L}
\newcommand{\Sigmatilde}{\widetilde \Sigma}

\newcommand{\caesar}{\textsc{caesar}\xspace}

\date{}

\subheader{CERN-PH-TH/2012-065\\
  LPN12-043\\
  OUTP-12-04P\\
}
\title{NLL+NNLO predictions for jet-veto efficiencies in Higgs-boson and Drell-Yan production}
\author[a]{Andrea Banfi}
\author[b,c,d]{Gavin P. Salam}
\author[e]{Giulia Zanderighi}

\affiliation[a]{Albert-Ludwigs-Universit\"at Freiburg, 
   Physikalisches Institut, D-79104 Freiburg, Germany.}
\affiliation[b]{CERN, PH-TH, CH-1211 Geneva 23, Switzerland.}
\affiliation[c]{Department of Physics, Princeton University, Princeton, NJ 08544, USA.}
\affiliation[d]{LPTHE; CNRS UMR 7589; UPMC Univ.\ Paris 6; Paris, France.}
\affiliation[e]{Rudolf Peierls Centre for Theoretical Physics,
  1 Keble Road, University of Oxford, UK.}

\emailAdd{andrea.banfi@physik.uni-freiburg.de, gavin.salam@cern.ch}

\emailAdd{g.zanderighi1@physics.ox.ac.uk}

\abstract{
  Using the technology of the \caesar approach to resummation, we
  examine the jet-veto efficiency in Higgs-boson and Drell-Yan
  production at hadron colliders and show that at next-to-leading
  logarithmic (NLL) accuracy the resummation reduces to just a Sudakov
  form factor.
  Matching with NNLO calculations results in stable predictions for the
  case of Drell-Yan production, but reveals substantial uncertainties in
  gluon-fusion Higgs production, connected in part with the poor behaviour of
  the perturbative series for the total cross section.
  We compare our results to those from \POWHEG with and without reweighting by
  \HqT, as used experimentally, and observe acceptable agreement.
  In an appendix we derive the part of the NNLL resummation
  corrections associated with the radius dependence of the jet
  algorithm.
}


\arxivnumber{1203.5773}

\begin{document}

\maketitle


\section{Introduction}
\label{sec:introduction}

One of the most active topics currently in particle physics is the
search for the Higgs
boson~\cite{ATLAS:2012si,Chatrchyan:2012tx,TEVNPH:2012ab}. 
This
scalar boson is the one particle of the standard model that remains to be
found, and its discovery would provide the most direct evidence to
date that the existence of a Higgs
field~\cite{Higgs:1964ia,Higgs:1964pj,Higgs:1966ev,Englert:1964et,Guralnik:1964eu}
is responsible for electroweak symmetry breaking.

Several production and decay channels are used to search for the Higgs
boson at the LHC and Tevatron, with varying sensitivities across the
range of Higgs-boson masses allowed in the standard model.
One channel with sensitivity over a broad range of masses is
gluon-fusion production followed by decay to a $W^+W^-$ pair, one of
which may be offshell.
For a Higgs mass of about $125\GeV$, as hinted at by current
data~\cite{ATLAS:2012si,Chatrchyan:2012tx}, this channel is one of
several that is expected to be clearly observable with forthcoming
data.
An accurate understanding of its cross section will therefore be
crucial in constraining the Higgs-boson couplings in the event of a
discovery.

Extensive work has been performed over the past decade to precisely
predict the cross sections and branching ratios for all the main
production and decay channels, as reviewed for example
in~\cite{Dittmaier:2011ti,Dittmaier:2012vm}.
In the case of Higgs boson production via gluon fusion with subsequent
decay to two $W$ bosons, a feature that is of particular importance in
discussing cross-section determinations is that it is customary for
the LHC experiments to separately treat events according to their
different jet multiplicities~\cite{ATLAS:2011aa,Chatrchyan:2012ty}.
This is because events with no jets are relatively free of backgrounds
other than $W^+W^-$ production, while events with one jet have
additional backgrounds from $t\bar t$ and Drell-Yan production, and
events with two jets or more also receive enhanced signal contributions
from vector-boson fusion production.

Of particular usefulness for the problem of evaluating cross sections
in gluon-fusion production with different jet multiplicities, are
fixed-order calculations that allow one to place arbitrary cuts on the
final state, up to next-to-next-to-leading order (NNLO) for inclusive
production~\cite{Anastasiou:2005qj,Grazzini:2008tf} and NLO for the
production in association with
one~\cite{deFlorian:1999zd,Ravindran:2002dc,Glosser:2002gm} or two
jets~\cite{Campbell:2006xx,Campbell:2010cz}. 
However, the transverse momentum $(p_t$) thresholds used for
identifying jets are usually well below the Higgs mass
($M_H$). This results in the appearance of logarithmically enhanced
terms at all orders of perturbation theory $\as^n \ln^{2n} M_H/p_t$,
which can spoil the convergence of fixed-order truncations of the
series.
Such problems arise also when examining other related final-state
observables, such as the Higgs-boson or Z-boson transverse momentum,
and the ``beam thrust''~\cite{Stewart:2009yx}, for which resummations
have been performed to next-to-next-to-leading-logarithmic accuracy
(NNLL) and combined with
NNLO~\cite{Bozzi:2003jy,Bozzi:2005wk,Berger:2010xi,Becher:2010tm},
or the transverse energy flow, resummed to
next-to-leading-logarithmic~\cite{Papaefstathiou:2010bw} accuracy.

Perhaps surprisingly, however, to date no resummation has been
performed for Higgs-related final-state definitions that involve jet
finding. This may be because jet related observables are less
inclusive than those resummed at high accuracy so far, and therefore
require a more detailed understanding of how multiple emissions affect
the value of the observable. For example in notable cases,
e.g. the $y_{23}$ jet resolution parameter of the exclusive
$k_t$-algorithm in $e^+e^-$ or hadronic 
collisions~\cite{Catani:1991hj,Catani:1993hr}, the existing NLL answer
has yet to be expressed in anything other than a numerical
form~\cite{Banfi:2001bz,Banfi:2004nk}.
Therefore, instead, the matched NNLL+NNLO results for other observables like the
Higgs $p_t$ have been used to reweight NLO hadronic event generators,
\MCatNLO~\cite{Frixione:2002ik} and \POWHEG~\cite{Nason:2004rx} (those
generators have been also used in standalone form).
Insofar as a jet veto differs from a Higgs-$p_t$ veto starting only
from order $\as^2$ relative to the Born process, the reweighting
procedure should provide reasonable predictions.
On the other hand, the modelling of inclusive Higgs production
processes in \MCatNLO or \POWHEG is such that the $\as^2$ difference
between the jet veto and Higgs $p_t$ veto is not correctly
included. Consequently some of the accuracy of the NNLL+NNLO
calculation is lost in the reweighting procedure.

The purpose of this paper is to examine the resummation, matching to
fixed order and resulting phenomenology directly for the jet veto
observable itself.
Because a jet definition is a less inclusive observable than the
Higgs-boson $p_t$ or the beam thrust, the resummation cannot be
immediately reduced to standard forms such as~\cite{Collins:1984kg}.
However, at NLL accuracy, to which we shall mostly limit ourselves here, one
may make use of the computer automated expert semi-analytical resummer
(\CAESAR)~\cite{caesar} to perform the resummation. 
The result turns out to be rather simple and straightforward also to
understand analytically.

For Higgs production, throughout this article, we will use the large
$m_{\rm top}$ approximation. This does not affect the resummation at our
accuracy but is relevant when we combine the resummation with fixed
order calculations.

In parallel with the Higgs-boson case, we will also examine the
jet-veto resummation, matching and phenomenology for Drell-Yan
production, which has been argued to provide a control case, even if,
as we shall see, the issues that arise in the Higgs-boson case are not
directly mirrored in Drell-Yan production.

\section{The jet-veto efficiency}
\label{sec:obs}
We consider the production of a Higgs or a Z boson accompanied by $N$
extra QCD partons $p_1,\ldots p_N$,
\begin{equation}
p p \to H + p_1+\ldots p_N\,,\quad {\rm and}\quad
p p \to Z + p_1+\ldots p_N\,. 
\label{eq:proc}
\end{equation}
A jet-veto condition is imposed by clustering the events into jets
using a suitable hadron-collider jet-definition (JD) and requiring
that the event has no jets with transverse momentum above a certain
threshold, typically in the range of $25-30\GeV$.
To define the jets, the LHC experiments usually use the anti-$k_t$
algorithm~\cite{Cacciari:2008gp}, which repeatedly merges the pair of
particles with smallest distance measure $d_{ij} =
\min(p_{ti}^{-2},p_{tj}^{-2}) \Delta R_{ij}^2/R^2$, unless there
exists a particle with a $d_{iB} = p_{ti}^{-2}$ value that is smaller,
in which case $i$ becomes a jet and is removed from the list of
particles.
Here $\Delta R_{ij}^2 = (y_i - y_j)^2 + (\phi_i - \phi_j)^2$, where
$y_i$ and $\phi_i$ are respectively the rapidity and azimuth of
particle $i$. The parameter $R$ sets the typical angular reach of the
jet definition and is often referred to as the jet radius.
After the clustering, one may choose to consider all jets, or
alternatively only those within some limited rapidity range, in
reflection of actual experimental acceptances.

The cross section with a jet-veto is defined as
\begin{equation}
  \Sigma(\ptjv) = \sum_{N} \int d\Phi_N
  \frac{d\sigma_N}{d\Phi_N} \,\Theta(\ptjv - \ptm^{(\text{JD})}(p_1,\ldots,p_N))\,, 
\label{eq:Sigma}
\end{equation}
where $p_{\rm t, j1}$ is the transverse momentum of the hardest
(highest $p_t$) of the jets found in the event, $d\sigma_N$ denotes
the partonic cross-section to produce a Higgs or a Z boson accompanied
by $N$ extra partons and $d\Phi_N$ is the corresponding phase space.

It is also useful to consider the jet-veto efficiency defined as
\begin{equation}
  \eff(\ptjv) \equiv 
  \frac{\Sigma(\ptjv)}{\sigma}\,, 
\label{eq:eff}
\end{equation}
where $\sigma$ is the total cross-section. 
The veto efficiency is of interest because it essentially encodes just
the information about the Sudakov suppression associated with
forbidding radiation of jets.
In contrast, the vetoed cross section mixes in also the physics that
determines the total cross section.
Thus, in the absence of a veto, $\ptjv = \infty$, the efficiency is
exactly $1$ and one can reliably discuss small departures from
$\eff = 1$ as $\ptjv$ is reduced.
In the vetoed cross section, $\Sigma(\ptjv)$, it is harder to
disentangle those effects from uncertainties on the total cross
section.
Later, we will argue that even for $\eff$ substantially below $1$,
it makes sense to treat the efficiency and total cross section as
independent quantities, and that the uncertainties that govern them
are relatively uncorrelated.

Each of $\Sigma(\ptjv)$, $\sigma$ and $\eff(\ptjv)$ has a fixed-order
perturbative expansion, which we write as
\begin{subequations}
  \begin{align}
    \label{eq:Sigma-expansion}
    \Sigma(\ptjv) &= \Sigma_{0}(\ptjv) + \Sigma_{1}(\ptjv) + \Sigma_{2}(\ptjv) + \ldots \,,\\
    \eff(\ptjv) &= \eff_{0}(\ptjv) + \eff_{1}(\ptjv) + \eff_{2}(\ptjv) + \ldots \,,\\
    \sigma &= \sigma_{0} + \sigma_{1} + \sigma_{2} + \ldots \,,
  \end{align}
\end{subequations}
where the index $i$ signifies that the contribution is proportional to
$\as^i$ relative to the Born term.
The properties $\Sigma_{0}(\ptjv) \equiv \sigma_{0}$ and $\eff_0 = 1$
follow from the fact that no jets are present at Born level.

\section{NLL Resummation}
\label{sec:resummation}

The jet-veto efficiency is nothing but the cumulative distribution of
the transverse momentum of the highest-$p_t$ jet, $\ptm$. In order to
have a dimensionless observable, we choose to divide it by
the corresponding boson mass, $M_B=M_Z$ or $M_H$,
\begin{equation}
V(\tilde p_1,\tilde p_2;k_1 \dots k_N) = \frac{\ptm(k_1 \dots k_N)}{M_B}\,.
\end{equation}
Here $\tilde p_1$ and $\tilde p_2$ denote the incoming partons entering
the hard scattering (after any initial state emission), $k_1\dots k_N$
are the momenta of the final-state QCD partons. The momentum of
the boson can be obtained using energy-momentum conservation. 

We present here a next-to-leading logarithmic (NLL) resummation of
the jet-veto efficiency, i.e.\ we resum all logarithms in $\ln
\eff(\ptjv)$ up to $\as^nL^{n}$, with $L\equiv \ln(M_B/\ptjv)$.
This observable is within the scope of the computer automated
resummation program \caesar~\cite{caesar}.
\caesar is a program that, given a computer subroutine for an
observable, automatically performs a numerical analysis of its
behaviour with respect to multiple soft and collinear emissions.
From this analysis it establishes whether the observable belongs to a
broad class for which it is able to perform NLL resummations, and if
so expresses the result for the resummation in terms of a ``master
formula'' together with various numerically determined input
parameters.
The physics framework that underlies \caesar can also be used in an
analytic context as we shall do, briefly, here.

The first element of the analysis of an observable is to establish its
dependence on the kinematics of a single soft and collinear emission. 
For the case of the jet veto, we simply have
\begin{equation}
  \label{eq:parametricform}
  V(\{{\tilde p}\}, k)=
  \frac{k_t}{M_B}\>,
\end{equation}
where $k_t$ is the transverse momentum of the emission with respect to
the beam. This expression holds exactly in all (hard, soft, collinear)
limits.

The next element is to establish whether the observable is {\em
  continuously global}~\cite{NG,DiscontGlobal}: essentially, the
observable should be non-zero for any emission with finite energy and
angle with respect to the beam; furthermore, the power-law of the
observable's dependence on the emission energy should be independent
of the emission angle.
Strictly speaking, the first part of this condition holds only if jets
are measured at 
all rapidities, whereas real experimental measurements have finite
acceptance. We shall however for now work assuming full acceptance
and return to the question in section~\ref{sec:compMC}.
Given the first part of the condition, the second part holds trivially
for the jet veto, and this ensures that the resummation is free of
``non-global'' logarithms and related terms.

A second condition was dubbed {\em recursive infrared and collinear
  (rIRC) safety}~\cite{caesar}, and concerns the observable's
behaviour in the presence of multiple emissions.
Essentially it involves two requirements: (a) if one scales all
emissions in some appropriately uniform manner towards the soft and
collinear limit, 
then the observable should scale in the same manner; and (b) if
one emission is soft relative to the others, and kept of fixed
relative softness while scaling all the emissions, then in the scaling
limit, the observable should not change if the soft emission is
removed.
Again, this condition holds trivially for the jet veto, as long as one
uses a standard ``inclusive'' longitudinally invariant infrared and
collinear-safe hadron collider jet algorithm
(e.g. $k_t$~\cite{Catani:1993hr,Kt-EllisSoper},
Cambridge/Aachen~\cite{Cam,Aachen},
anti-$k_t$~\cite{Cacciari:2008gp}).
It ensures that double-logarithmic terms exponentiate and that one
can use an independent-emission type approximation in evaluating NLL
terms.

The master formula in the \caesar approach then tells us that the NLL
resummed jet-vetoed cross section is given by
\begin{subequations}
  \label{eq:Master}
  \begin{multline}
    \Sigma_{\rm NLL}(\ptjv) = \int dx_1 dx_2 \, f\!\left(x_1, \ptjv\right)
    f\!\left(x_2, \ptjv\right) 
    \delta(x_1 x_2 s - M_B^2)\, |{\cal M}_B|^2 
    \\ \cdot 
    \cF(R_B') e^{-R_B\left(\ptjv\right)}\,,
  \end{multline}
  where $|{\cal M}_B|^2$ is the full tree-level matrix element squared to
  produce a Z or a Higgs boson including the flux, and spin- and
  colour-averages; $e^{-R_B}$ is a Sudakov form factor, with $R_B$ a
  double-logarithmic function given 
  respectively for the $Z$ and Higgs-boson cases by
  \begin{align}
    R_Z\left(\ptjv\right) 
    &= 2C_F \int_{\ptjv^2}^{M_Z^2}
    \frac{dk_t^2}{k_t^2}\frac{\asCMW(k_t)}{\pi}
    \left(\ln \frac{M_Z}{k_t}-\frac{3}{4}\right)\,,
    \label{eq:RZ}
    \\
    R_H\left(\ptjv\right) 
    &= 2C_A \int_{\ptjv^2}^{M_H^2}
    \frac{dk_t^2}{k_t^2}\frac{\asCMW(k_t)}{\pi}
    \left(\ln \frac{M_H}{k_t}-\frac{11\CA - 4\TR\nf}{12\CA}\right)  \,,
    \label{eq:RH}
  \end{align}
  with $\asCMW(k_t)$ the coupling constant in the CMW
  scheme~\cite{Catani:1990rr}, evaluated at scale $k_t$.
\end{subequations}
The quantity $R_B'$ is equal to $dR_B/d\ln({M_B}/{\ptjv})$ and to NLL
accuracy it is simply $4 {C \as(\ptjv)}/{\pi} \ln
({M_B}/{\ptjv})$. It enters in the
function $\cF(R_B')$, which provides the single-logarithmic
contributions associated with the observable's dependence on multiple
emissions.
As derived in \cite{caesar} (and references therein), $\cF(R_B')$ is
given by
\begin{multline}
  \label{eq:cF-with-vto0-limit-anyn}
  \cF (R_B') = \lim_{\epsilon\to0}
  \epsilon^{R_B'}
    \sum_{m=0}^{\infty} \frac{1}{m!}
  \left( \prod_{i=1}^{m} \sum_{\ell_i=1}^2
    \frac{R_B'}{2}\int_\epsilon^{\infty} \frac{d\zeta_i}{\zeta_i} \int_0^1 d\xi_i
    \int_0^{2\pi} \frac{d\phi_i}{2\pi}
  \right) 
  \times \\ \times 
  \Theta\left(1 - \lim_{\bar v\to0} \frac{V(\{\tilde
      p\},\kappa_1(\zeta_1 \bar v), \ldots,
      \kappa_{m}(\zeta_{m} \bar v))}{\bar v}\right),
\end{multline}
where the $\kappa_i(\zeta_i \bar v)$ are parametrised momenta with
azimuthal angle $\phi_i$, rapidity $y_i = (-1)^{\ell_i} \xi_i \ln
1/{\bar v}$ and $k_{ti} = \zeta_i \bar v M_B$, with the hard incoming
momenta $\{\tilde p\}$ adjusted to ensure momentum conservation.
In evaluating $\cF$, a key point is that in the limit $\bar v \to 0$
all emissions become widely separated in rapidity. Consequently the
jet algorithm, independently of the jet radius $R$ and whether it's of
the SISCone~\cite{Salam:2007xv} or
generalised-$k_t$~\cite{Catani:1993hr,Kt-EllisSoper,Cam,Aachen,Cacciari:2008gp}
type, clusters each particle into a separate jet.
The leading jet will then be given by the hardest of all the
emissions, so that 
\begin{equation}
  \label{eq:result-for-observable}
  \lim_{\bar v\to0} \frac{V(\{\tilde
    p\},\kappa_1(\zeta_1 \bar v), \ldots,
    \kappa_{m}(\zeta_{m} \bar v))}{\bar v} =  \max(\zeta_1, \ldots,\zeta_n)\,.
\end{equation}
It is then straightforward to show that $\cF(R_B') = 1$. This can be
verified also numerically with \caesar.
The result $\cF(R_B') = 1$ has been found before also for the
resummation of the jet-resolution parameter in the $e^+e^-$ Cambridge
algorithm~\cite{Banfi:2001bz}.

The feature of the jet veto observable that causes $\cF(R_B') = 1$ has
consequences for the resummation beyond NLL. In particular certain key
observable-dependent pieces of the NNLL resummation were outlined in
the appendices of \cite{caesar}. These are discussed further in
appendix~\ref{sec:NNLL-R-dependence}. For the remainder of the main
part of the article, however, we restrict our attention to NLL
resummation.

In what follows below, rather than using Eq.~(\ref{eq:Master})
directly as written, we shall reduce it to a form where the exponent
involves just LL and NLL terms, expressed in terms of a logarithm $L =
\ln Q/ \ptjv$, with $Q$ the hard scale with respect to which the
resummation is defined, as well as the $\MSbar$ coupling $\as(\mu_R)$
at a given renormalisation scale $\mu_R$ and PDFs with a hard
factorisation scale $\mu_F$.
It is to be understood that $Q$, $\mu_R$ and $\mu_F$ will all be taken
of order $M_B$.
Varying them in the neighbourhood of $M_B$ will allow us to probe the
impact of subleading terms.
In practice, this is accomplished by expressing the CMW coupling in
terms of the $\MSbar$ coupling
\begin{equation}
  \label{eq:asCMW}
  \asCMW = \as\left(1+K \frac{\as}{2 \pi}\right)\,,\qquad
K = C_A \left(\frac{67}{18}-\frac{\pi^2}{6}\right) - \frac{5}{9}
n_f\,,
\end{equation}
and using for $\as(k_t)$ the following expression
\begin{equation}
  \label{eq:askt}
  \as(k_t) = \frac{\as(\mu_R)}{1-2\rho}
  \left(1-\as(\mu_R)\frac{\beta_1}{\beta_0}
    \frac{\ln(1-2 \rho)}{1-2 \rho}\right)\,,
  \qquad \rho \equiv \as(\mu_R)\beta_0\ln\frac{\mu_R}{k_t}\,,
\end{equation}
with 
\begin{equation}
\beta_0 = \frac{11 C_A - 2 n_f}{12\pi}\,\qquad \beta_1 = \frac{17
  C_A^2 - 5 C_A n_f - 3 C_F n_f}{24\pi^2}\,.
\end{equation}
A closed result for the NLL resummed result can then be obtained by
substituting Eqs.~(\ref{eq:asCMW},\ref{eq:askt}) into
Eq.~(\ref{eq:Master}):
\begin{multline}
  \label{eq:Sigma-with-g1-g2}
  \Sigma_{\rm NLL}(\ptjv) = \int dx_1 dx_2 \, f\!\left(x_1, \mu_F e^{-L}
    \right)
  f\!\left(x_2, \mu_F e^{-L}\right) 
  \delta(x_1 x_2 s - M_B^2)\, |{\cal M}_B|^2 
  \\ \cdot 
  \exp\left[ L g_1(\as(\mu_R) L) + g_2(\as(\mu_R)L)\right]\,,
\end{multline}
with
\begin{subequations}
  \label{eq:g1g2}
  \begin{align}
    \label{eq:g1}
    g_1(\as L) &= \frac{2 C}{\pi \beta_0} \left(1 + \frac{\ln(1
      - 2\lambda)}{2\lambda}\right)\,,
    \\
    \label{eq:g2}
    g_2(\as L) &= \frac{2 C}{\pi \beta_0} \left[
      -\left(\frac{K}{4 \pi\beta_0} + \ln \frac{\mu_R}{M_B} \right)
      \left(
        \ln\left(1-2\lambda\right)
        +\frac{2\lambda}{1-2\lambda}
      \right) + B\ln(1-2\lambda) 
    \right.\\\nonumber
    &  \qquad \qquad\quad \left.+ \frac{\beta_1}{2\beta_0^2} 
      \left(
        \frac12 \ln^2\left(1-2\lambda\right)
        +\frac{\ln\left(1-2\lambda\right)
          +2\lambda}{1-2\lambda}
      \right)-\ln \frac{M_B}{Q}\frac{2\lambda}{1-2 \lambda}
      \right]\,,
  \end{align}
\end{subequations}
where, for Higgs production $C=C_A$ and $B=-(11 C_A-2 n_f)/(12 C_A)$,
whilst for Drell-Yan production, $C=C_F$ and $B=-3/4$, and
$\lambda = \alpha_s(\mu_R) \beta_0 L$.
This form has the property that in the exponent there are
\emph{only} LL and NLL terms, whereas Eq.~(\ref{eq:Master})
effectively also contained some higher-order contributions.
Note also that the use in Eq.~(\ref{eq:Sigma-with-g1-g2}) of $\mu_F
e^{-L}$ rather than $\ptjv$ in (\ref{eq:Master}) corresponds to a NNLL
difference.

One advantage of formulating the resummed result in terms of the
variable $L$ is that it is then straightforward to adapt the
resummation so as to ensure a sensible behaviour even for
$\ptjv \gtrsim M_B$.
This is done by ``modifying'' the logarithm, 
\begin{equation}
  L \to \Ltilde \equiv
  \frac{1}{p}\ln \left(\left(\frac{Q}{\ptjv}\right)^p
    -\left(\frac{Q}{\ptM}\right)^p+1\right)\,,
\label{eq:Ldef}
\end{equation}
where $\ptM$ is the largest physically accessible jet momentum.
For $Q \ll \ptjv \ll \ptM$, $\Ltilde \sim (Q/\ptjv)^p/p$ and thus $p$
can be chosen so as to ensure that resummation effects vanish quickly
above the scale $Q$.
In practice we choose $p=5$.
This should help limit artifacts that could arise with the matching
procedure introduced later on in this article.

The above way of formulating the resummation follows in the footsteps
of~\cite{CTTW}, and includes refinements proposed
in~\cite{DISresum}. One of these refinements is that we allow $Q \neq
M_B$, which accounts for the $\order{1}$ additive freedom in defining
$L$ and the point where the resummation sets in. It plays an important
role in uncertainty estimates.

For completeness we explicitly write the final form we use for the
resummation cross section as $\Sigmatilde$, which is simply
Eq.~(\ref{eq:Sigma-with-g1-g2}) with $L$ replaced by $\Ltilde$:
\begin{multline}
  \label{eq:Sigmatilde-with-g1-g2}
  {\Sigmatilde}_{\rm NLL}(\ptjv) = \int dx_1 dx_2 \,
  f\!\left(x_1, \mu_F e^{-\Ltilde}
    \right)
  f\!\left(x_2, \mu_F e^{-\Ltilde}\right) 
  \delta(x_1 x_2 s - M_B^2)\, |{\cal M}_B|^2 
  \\ \cdot 
  \exp\left[ \Ltilde g_1(\as(\mu_R) \Ltilde) + g_2(\as(\mu_R)\Ltilde)\right]\,.
\end{multline}

\begin{figure}
  \centering
  \includegraphics[width=0.49\tw]{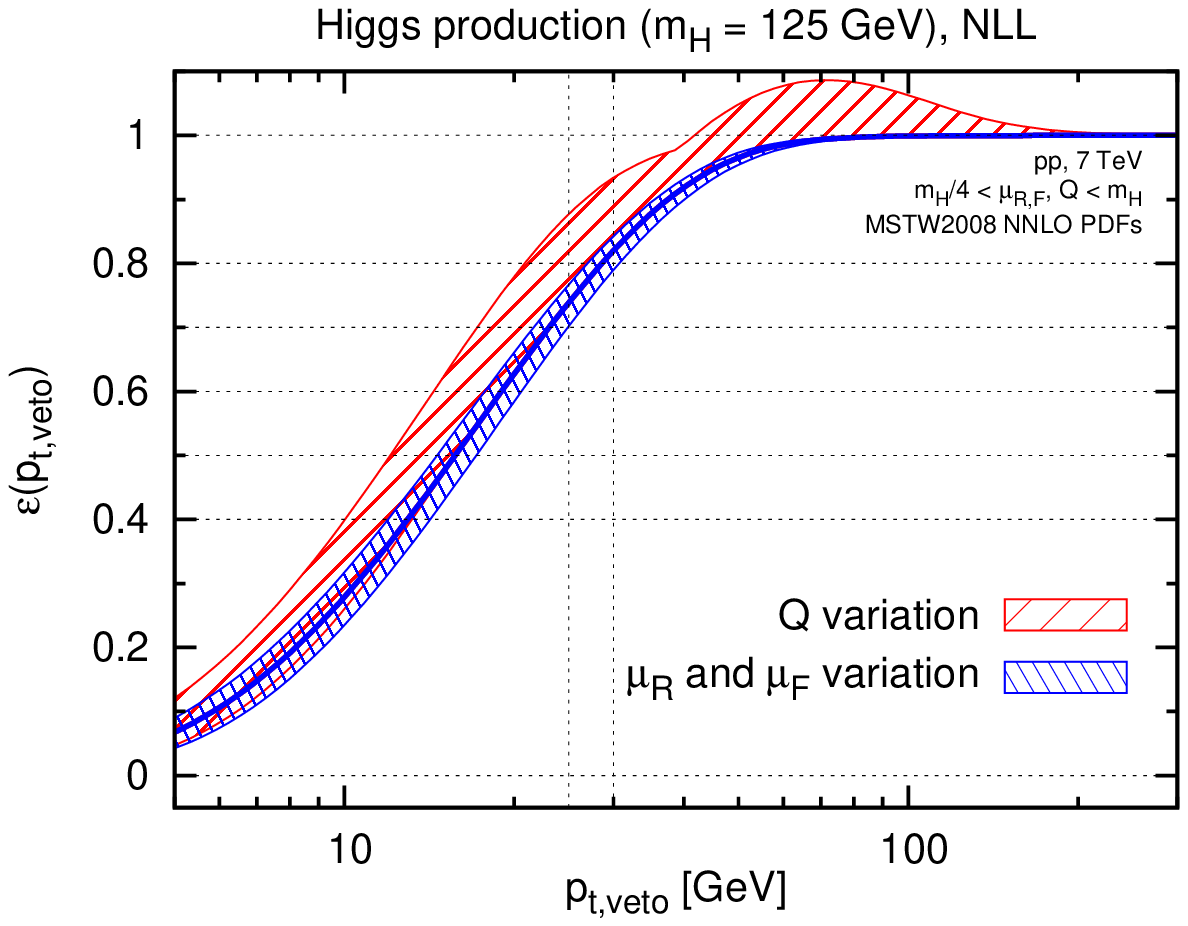}
  \includegraphics[width=0.49\tw]{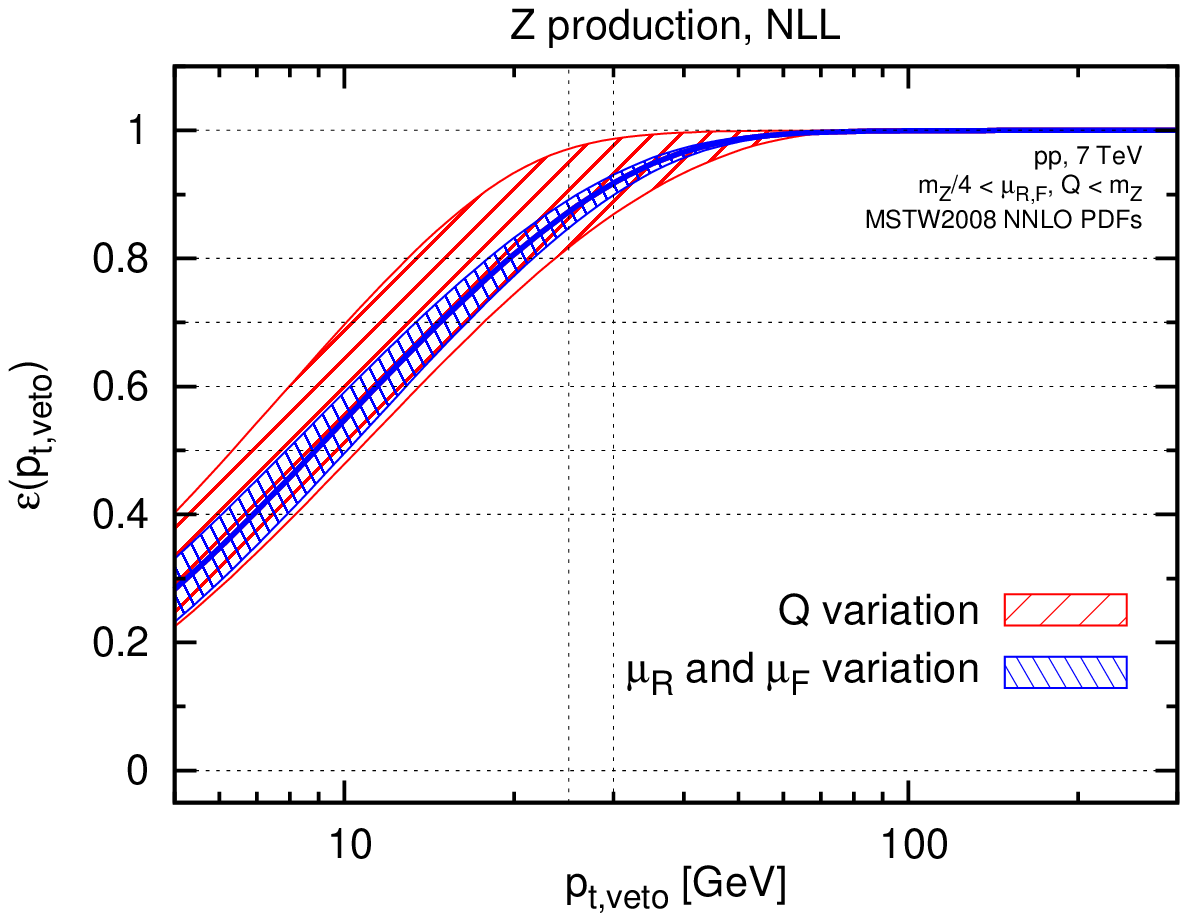}
  \caption{NLL resummed result for the jet-veto efficiency in Higgs
    (left) and $Z$-boson (right) production. The predictions are for
    $pp$ collisions at a centre of mass energy of $7\TeV$ and use
    MSTW2008 NNLO PDFs~\cite{Martin:2009iq}. For Higgs production we
    use the large $m_{\rm top}$ approximation.}
  \label{fig:pure-resummation}
\end{figure}

Figure~\ref{fig:pure-resummation} illustrates the resummed results for the
Higgs-boson and $Z$-production cases.
It shows the efficiency defined as
\begin{equation}
  \label{eq:eff-NLL}
  \eff_{\rm NLL}(\ptjv) = 
     \frac{ \Sigmatilde_{\rm NLL}(\ptjv)}{ \Sigmatilde_{\rm NLL}(\ptM)}\,.
\end{equation}
The vertical dashed lines mark the values of the jet-veto thresholds
used by ATLAS and CMS experiments, of 25 GeV and 30 GeV, respectively.
The thick solid (blue) line corresponds to our central scale choice
$\mu_R = \mu_F = Q = M_B/2$.
The finely hashed (blue) band around it corresponds to the envelope of
results obtained when examining an independent variation of the
renormalisation and factorisation scales with a constraint on their
maximal ratio:
\begin{equation}
  \label{eq:RF-scale-choices}
  \mu_R= \left\{\frac{M_B}4, \frac{M_B}2, M_B\right\},\qquad
  \mu_F = \left\{\frac{M_B}4, \frac{M_B}2, M_B\right\},\qquad
  \frac12 \,\le\, \frac{\mu_R}{\mu_F} \,\le\, 2\,.
\end{equation}
The resulting uncertainty is relatively modest. 
More significant is the uncertainty associated with the
variation of $Q$, shown by the widely-hashed (red) band, corresponding
to the envelope of the choices $Q = \{M_B/4, M_B/2, M_B\}$ (all with $\mu_R
= \mu_F = M_B/2$).
This variation probes the impact of unknown NNLL terms. 
In the case of Higgs production, one sees that the band even goes
almost $10\%$ above $1$. This kind of behaviour is common
in NLL resummation and can be traced back to the $B$
term in Eq.~(\ref{eq:g2}) or equivalently the second, negative term in
the round brackets of Eq.~(\ref{eq:RH}).
The problem is less severe in the $Z$-production case, because of a
cancellation due to PDF effects.
These kinds of artifacts are one of the reasons why it is important to
``match'' resummations with fixed order predictions, especially in the
region of moderate to large $p_t$ vetoes. 
In the next section we will therefore consider the structure of the
fixed-order cross sections, and subsequently proceed to introduce our
matching prescriptions and to examine the results.

\section{Jet-veto at fixed order}
\label{sec:perturbative-results}

The state-of-the-art of fixed-order predictions for fully differential
differential partonic Higgs-boson and Z-boson cross sections is NNLO,
i.e.\ the calculation of $\Sigma_2(\ptjv)$ and $\sigma_2$, with tools
like \FEHIP\cite{Anastasiou:2005qj} and \HNNLO~\cite{Grazzini:2008tf}
for Higgs productions, and \FEWZ\cite{Gavin:2010az} and
\DYNNLO\cite{Catani:2009sm} for Z production.
For the purpose of determining the jet-veto cross section, it is
however also possible (and sometimes numerically cheaper) to compute
only $\sigma_2$ with these NNLO tools (or from the inclusive
results~\cite{Harlander:2002wh,Anastasiou:2002yz,Ravindran:2003um,Hamberg:1990np}),
and obtain $\Sigma_1(\ptjv)$ and $\Sigma_2(\ptjv)$ from the relation
\begin{equation}
  \label{eq:Sigma-from-diff}
  \Sigma_{i}(\ptjv) = \sigma_{i} + \bar \Sigma_i(\ptjv), \qquad 
\bar \Sigma_i(\ptjv) = - \int_{\ptjv}^\infty\, d\pt \frac{d\Sigma_{i}(\pt)}{d\pt}\,.
\end{equation}
The differential distributions $d\bar \Sigma_{1}/d\pt$ and
$d\bar \Sigma_{2}/d\pt$ can be computed from the 
boson+jet cross sections at LO and NLO respectively, e.g. using
MCFM~\cite{Campbell:2006xx,Campbell:2010cz, Campbell:2002tg}.
We recall that, throughout, we use the large $m_{\rm top}$ approximation
for Higgs production.

\subsection{Prescriptions for the efficiency}

There is little ambiguity in the definition of the fixed order results
for the total and jet-vetoed cross-sections, with the only freedom
being, as usual, in the choice or renormalisation and factorisation
scale. 
However, given the expressions of $\Sigma$ and $\sigma$ at a given
perturbative order, there is some additional freedom in the way one
computes the jet-veto efficiency. For instance, at NNLO the efficiency
can be defined as 
\begin{subequations}
  \begin{equation}
    \eff^{(a)}(\ptjv) \equiv \frac{\Sigma_0(\ptjv)+\Sigma_1(\ptjv)+\Sigma_2(\ptjv)}{\sigma_0+\sigma_1+\sigma_2}\,, 
    \label{eq:NNLOa}
  \end{equation}
  but the following expressions are equally valid at NNLO,
  \begin{align}
    \label{eq:NNLOb}
    \eff^{(b)}(\ptjv) &\equiv
    \frac{\Sigma_0(\ptjv)+\Sigma_1(\ptjv)+\bar
      \Sigma_2(\ptjv)}{\sigma_0+\sigma_1}\,, \\
    \eff^{(c)}(\ptjv) &\equiv 1 +\frac{\bar \Sigma_1(\ptjv)}{\sigma_0}+
    \left(\frac{\bar
        \Sigma_2(\ptjv)}{\sigma_0}-\frac{\sigma_1}{\sigma_0^2}\bar
      \Sigma_1(\ptjv)\right) \,,
    \label{eq:NNLOc}
  \end{align}
\end{subequations}
since they differ relative to Eq.~(\ref{eq:NNLOa}) only by terms
$\cO{\as^3}$, which are not under control.

Option $(a)$ is the most widely used, and may appear at
first sight to be the most natural, since one keeps as
many terms as possible both in the numerator and denominator.
However, option $(b)$ can be motivated as follows: since the zeroth
order term of $\epsilon(\ptjv)$ is equal to $1$, it is really only
$1-\epsilon(\ptjv)$ that has a non-trivial perturbative series, given
by the ratio of the inclusive 1-jet cross section above $\ptjv$,
$\sigma_{\text{1-jet}}^\text{NLO}(\ptjv)$, to the total cross section.
Insofar as the 1-jet cross section is known only to NLO, in taking the
ratio to the total cross section one can argue that one should also
use NLO for the latter, i.e.\
\begin{equation}
  \label{eq:NNLOb-motivation}
  \epsilon(\ptjv) = 1 -
  \frac{\sigma_{\text{1-jet}}^\text{NLO}(\ptjv)}{\sigma_0 + \sigma_1}\,.
\end{equation}
It is straightforward to verify that this then leads to
Eq.~(\ref{eq:NNLOb}).
This procedure also coincides with the one adopted in event-shape studies
in DIS and hadron-hadron collisions ($\sigma_2$ is not even known in
the latter case).
Option $(c)$ is also well motivated, since it is a strict fixed order
expansion of the ratio, so no uncontrolled terms beyond NNLO are
included. This is the prescription that is usually adopted in $\ee$
event-shape and jet-rate studies.

While other possibilities are also equally valid, the above three
schemes capture a substantial part of the freedom that one has in
writing the series.
The size of the differences between them is one way to estimate the
associated theoretical uncertainty and goes beyond the usual variation
of scales.

\subsection{Numerical results}
\label{sec:NNLO-numerical}
\begin{figure}[tbp]
\includegraphics[width=0.49\textwidth]{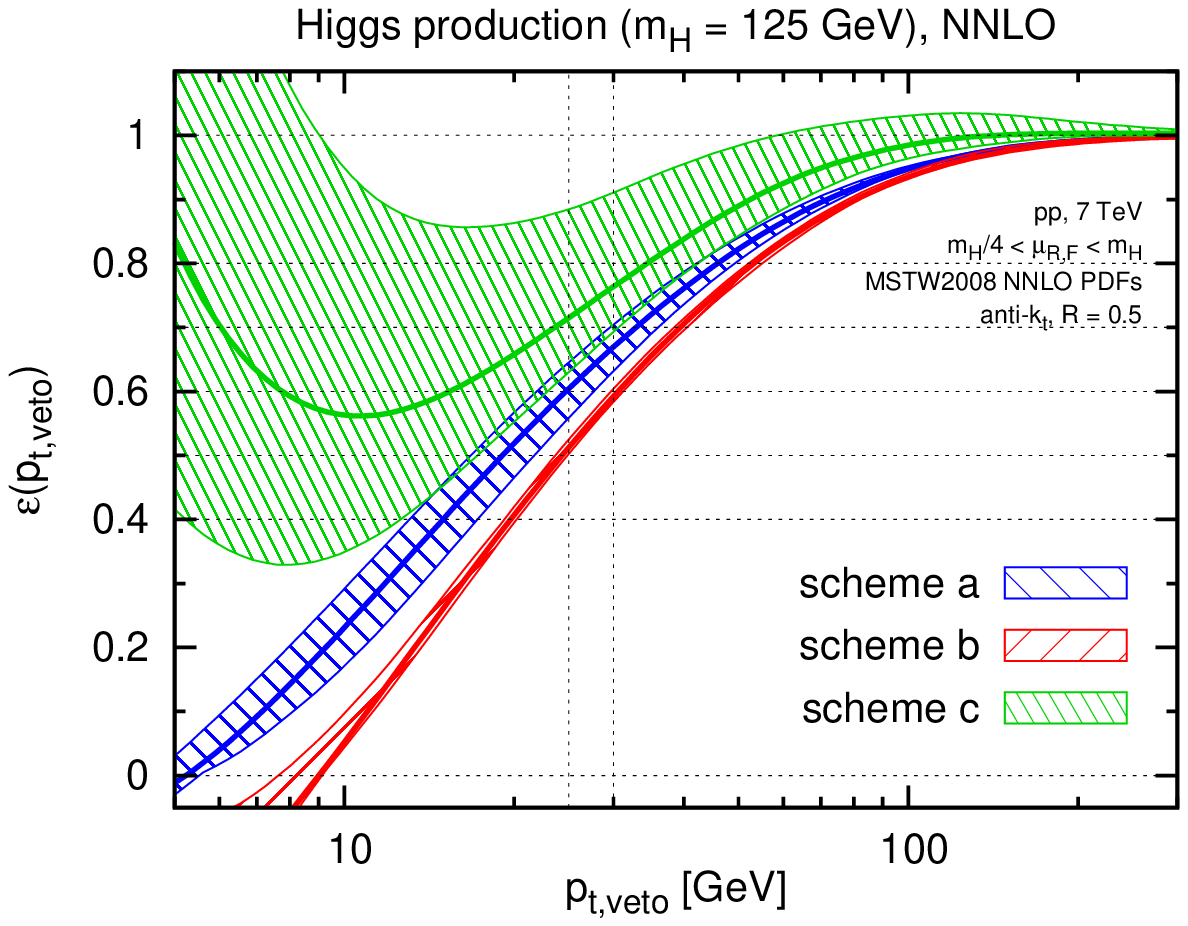}
\includegraphics[width=0.49\textwidth]{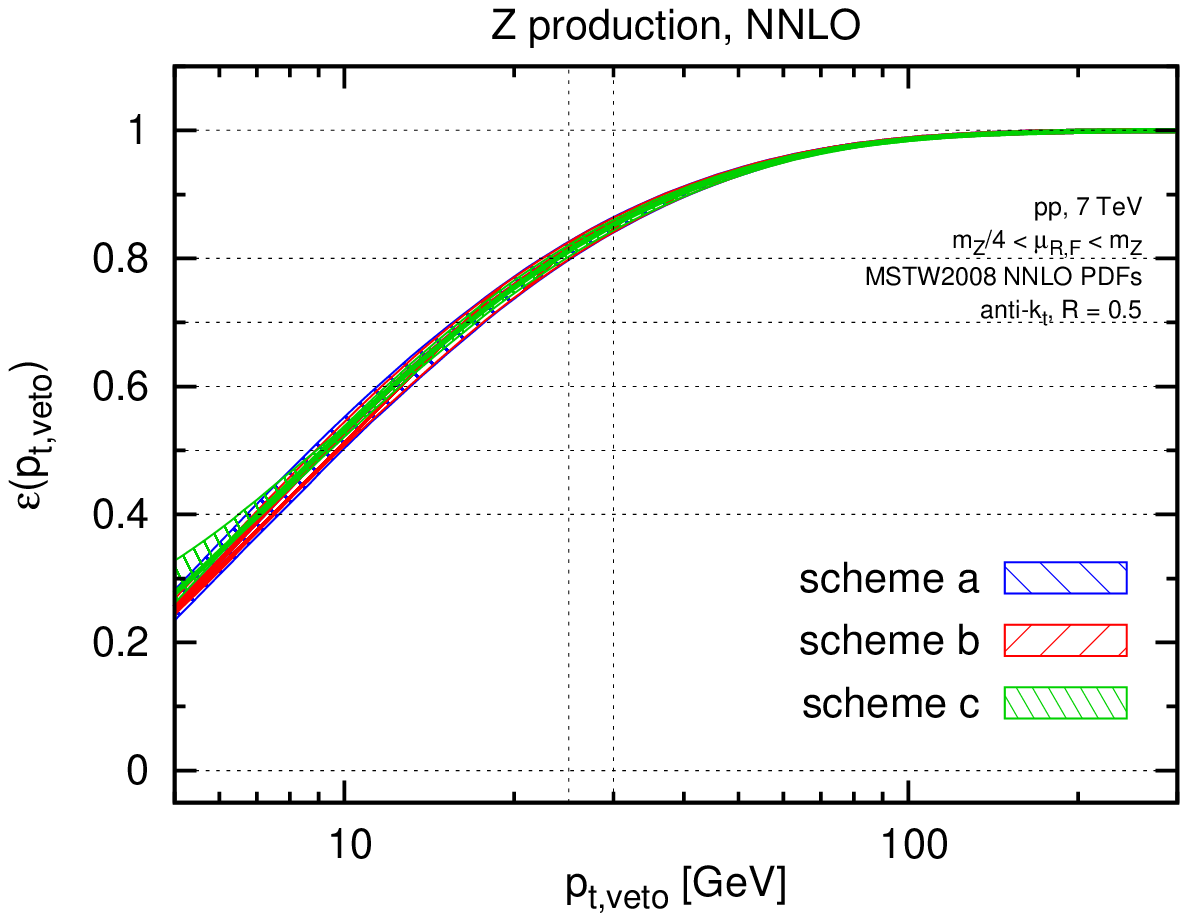}
\caption{Jet-veto efficiency for Higgs (left) and $Z$-boson production
  (right) using three different prescriptions for the NNLO expansion,
  see Eqs.(\ref{eq:NNLOa}--\ref{eq:NNLOc}). 
  For each prescription, the thick solid line corresponds to the result obtained with $\mu_R =
  \mu_F = M_{H/Z}/2$, while the
  band shows the scale uncertainty as obtained with the
  choices of Eq.~(\ref{eq:RF-scale-choices}).
}
\label{fig:ptjv3NNLO}
\end{figure} 

Figure~\ref{fig:ptjv3NNLO} shows the NNLO results for the jet-veto
efficiency in the 3 schemes discussed above.
Each scheme is displayed as a band corresponding to the envelope of
the scale variations as in Eq.~(\ref{eq:RF-scale-choices}), together
with a solid line for the prediction with the central scale choice.

In the case of Higgs production (left-hand plot) the bands barely
overlap and, in the region of interest, $\ptjv\sim 25-30$~GeV,
the three predictions differ considerably, with the bands spanning the
range from 0.6 to 0.9 and the central values from $0.6$ to $0.77$
(for $\ptjv = 30$~GeV).
For yet smaller $\ptjv$ values one sees a complete breakdown of the
fixed-order predictions, with the efficiencies ranging from below $0$
to above $1$.
Curiously, scheme-c misbehaves also at large $\ptjv \sim M_H$, where it
leads to an efficiency larger than $1$.

In contrast, for $Z$ production (right-hand plot), the situation is
much more stable, with all three schemes overlapping and remaining
sensible-looking down even to low $\ptjv$.

The large differences between the Higgs and $Z$-boson cases call for
an explanation.
One consideration is that the $gg$ initial state that is relevant for
Higgs production implies $C_A=3$ colour factors rather than $C_F=\frac43$ colour
factors. 
This inevitably worsens the perturbative convergence.
The larger colour factor is however not the only culprit. 
Table~\ref{tab:sigma} shows the total cross sections for Higgs and $Z$
production calculated at different orders, as well as the coefficients
$c_i$ of the perturbative expansion normalised to $\sigma_0$ at the
central scale choice, i.e.\ $c_i = \sigma_i /(\sigma_0 \as^i)$ such
that the cross section reads
\begin{equation}
  \label{eq:sigma-ci}
  \sigma = \sigma_0 (1 + c_1 \as + c_2 \as^2 + \ldots)\,. 
\end{equation}
The difference between schemes $a$ and $b$ is in the choice 
of whether the $1$-jet cross section is normalised to the NLO or NNLO
total cross section. The NNLO term brings a correction of order $20\%$
for Higgs production and this is reflected in the efficiencies;
for $Z$ production, $\sigma_2$ is essentially zero and so
schemes $a$ and $b$ give identical results.
As concerns scheme $c$, one is additionally affected by the very large
size of $\sigma_1$ in the Higgs
case~\cite{Dawson:1990zj,Djouadi:1991tka}. 

Table~\ref{tab:sigmawithveto} shows corresponding cross section results with a
jet-veto of $30 \GeV$.
The Z-production case maintains its reasonable convergence.
Interestingly, the vetoed Higgs cross section now also appears to have
a convergence that is significantly improved with respect to the total
cross section.
As has been argued by others~\cite{Stewart:2011cf}, this apparent
improvement in convergence can be interpreted as an artifact of
cancellations between two independent classes of large corrections:
those that lead to the poor convergence of the total cross section and
those associated with the Sudakov suppression for small jet vetoes
(enhanced compared to the Z case due to the larger colour factors for
Higgs production).
The presence of such artificial cancellations in the cross section is
one of the reasons that we prefer to consider the jet veto efficiency.
Furthermore, as we have seen, the use of the jet veto efficiency
provides additional handles to estimate the perturbative uncertainty
through the three schemes discussed above.

\begin{table}
\begin{center}
\begin{tabular}{ccccccc}
  \toprule
  &LO& NLO& NNLO&& $c_1$ &$c_2$ \\\midrule
 H [pb] & $5.40^{+1.43}_{-1.02}$ & $11.96^{+2.63}_{-1.95}$ &
  $14.7^{+1.2}_{-1.4}$ & & 9.78 & 33.5 \\[5pt] 
  Z [nb] & $22.85^{+2.07}_{-2.40}$ & $28.6^{+0.8}_{-1.2}$ &
  $28.6^{+0.4}_{-0.4}$ & & 1.94 & 0.0 \\
  \bottomrule
\end{tabular}
\end{center}
\caption{Cross-sections for Z and Higgs (large $m_{\text{top}}$
  approximation) production at various
  orders in perturbation theory. The central value corresponds to the
  default scale, $\mu_R=\mu_F=M_B/2$, $M_B = M_Z$ or $M_H$, the error denotes the scale variation when
  $\mu_R$ and $\mu_F$  are varied independently by a factor two around 
  the central value, with the constraint that $\mu_R/2 \le \mu_F \le 2
  \mu_R$, i.e.\ Eq.~(\ref{eq:RF-scale-choices}). 
  Also shown are the expansion coefficients $c_i$ of the cross-sections
  for the central scale $\mu_0 = M_B/2$, $\sigma_i = \sigma_0 c_i
  \as^i(\mu_0)$.
  The results were obtained using the \HNNLO~\cite{Grazzini:2008tf}
  and \DYNNLO\cite{Catani:2009sm} programs, with 
  MSTW2008 NNLO PDFs and $\as(M_Z) =0.11707$, and  a centre-of-mass
  energy of $7 \TeV$.
}
\label{tab:sigma}
\end{table}

\begin{table}
\begin{center}
\begin{tabular}{ccccccc}
  \toprule
  &LO& NLO& NNLO && $c_1$ &$c_2$ \\
  \midrule
  H [pb] & $5.40^{+1.43}_{-1.02}$ & $8.99^{+1.33}_{-1.16}$ &
  $9.88^{+0.27}_{-0.55}$ && 5.36&  10.6\\[5pt] 
  Z [nb] & $22.85^{+2.07}_{-2.40}$ & $25.54^{+0.64}_{-1.02}$ & $24.5^{+0.6}_{-0.7}$ &&$0.90$  &$-2.7$  \\
  \bottomrule
\end{tabular}
\end{center}
\caption{Same as Tab.\ref{tab:sigma} but with a
  jet-veto of $30\GeV$, based on the anti-$k_t$ algorithm with R=0.5. The
  LO cross-section is equal to that without a veto.
  The results have been determined by subtracting $H+1$-jet and
  $Z+1$-jet cross sections, obtained with MCFM, 
from the inclusive cross
  sections obtained with \DYNNLO and \HNNLO, though we could equally
  well have used \DYNNLO and \HNNLO directly.
 }
\label{tab:sigmawithveto}
\end{table}

Given that we have indicated that the uncertainties on the Higgs
jet-veto efficiency are associated with the poor convergence of the
total Higgs cross section, some comments are due concerning
discussions in the literature on our knowledge of the total Higgs
cross section.
On one hand there is work that aims to account for threshold and other
enhanced terms beyond
NNLO~\cite{Catani:2003zt,Kidonakis:2007ww,Ahrens:2008qu,Ahrens:2008nc}, 
while other work has suggested that there is a need to revisit, and
perhaps be more conservative in, estimates of the uncertainties on the
total cross section~\cite{Baglio:2010ae}.
Regarding improvements of the total cross section, while in general
these have the potential to be highly valuable, a more accurate
evaluation of the total cross section is, we believe, likely to
provide little improvement for the jet-veto efficiency: 
indeed, our results indicate that there are potentially large missing
$\order{\as^3}$ corrections to the jet veto efficiency and it is only
through an evaluation of the NNLO H+1-jet rate that these could be
fully constrained.



\section{Prescription for estimating total uncertainties}
\label{sec:uncertainty-prescription}

From now onwards, for the purpose of quoting final efficiency
estimates, we will need a prescription for combining different sources
of uncertainty.
We shall adopt the envelope method, used for example in
\cite{Jones:2003yv} as well as in many other works on resummation.
This method takes the envelope of the curves obtained from each of
several different sources of uncertainty estimate. The logic of the
scheme is that to avoid double counting uncertainties, at most one
source should be probed at a time.
Thus here at fixed order we would take the envelope that results from
the scale variation band from one main efficiency prescription (scheme
$a$) and the central values of each of the two alternative efficiency
prescriptions ($b$ and $c$).
As has been found in \cite{LesHouches}, this leads to an uncertainty
estimate that is quite close to the alternative fixed-order
prescription proposed in \cite{Stewart:2011cf}, which varies the scale
choice independently in zero, one and two-jet bins.

In addition to estimating the uncertainty on the veto efficiency,
it is also important to be able to estimate the uncertainty on the
final vetoed cross section.
The prescription that we propose is to treat the efficiency
uncertainty as being uncorrelated with the uncertainty on the total
cross section.
This can be justified on the following grounds: the total cross
section uncertainty is associated with our ignorance of the NNNLO
corrections to its perturbative series.
On the other hand the uncertainties in the efficiency are related to
higher-order Sudakov type terms and to the way in which we treat the
known NNLO cross-section contributions in calculating the efficiency.
These are sufficiently different in origin that we believe that it is
reasonable to treat them as uncorrelated.

\section{Matching NLL and NNLO results}
\label{sec:matchNLLNLO}

To obtain predictions for the jet veto efficiency that include the
advantages of both resummed and fixed order results, we combine them
with the help of a matching procedure. 
Below, we first briefly recall the various requirements that
matching should satisfy, then list the matching prescriptions that we
adopt, and finally discuss the matched results.

\subsection{Matching prescriptions}

As is well-known there are various ways in which one can match resummed
and fixed order calculations. Since we match the resummation to NNLO
exact results, the  matching procedure should satisfy the
following three requirements:
\begin{itemize}
\item[1.] The matched result should be correct up to NLL terms in the
  exponent and the expanded matched result should be correct up to and
  including $\cO{\as^n L^{2n-2}}$ terms. 
\item[2.] The expanded matched result should coincide with
  the fixed order result up to and including the NNLO terms.
\item[3.] The jet-veto efficiency should tend to one at the maximum
  allowed jet transverse momentum $\ptM$, and the corresponding
  differential distribution ${d\eff(\ptjv)}/{d\ptjv}$ should vanish at
  that point 
  \begin{equation} 
    \label{eq:matching-at-kinematic-limit}
    \eff(\ptjv) = 1\quad{\rm and}\quad  
    \frac{d\eff(\ptjv)}{d\ptjv} = 0\quad{\rm for}\quad \ptjv = \ptM\,. 
  \end{equation} 
  The use of the modified logarithm $\Ltilde$, defined in
  Eq.~(\ref{eq:Ldef}) makes it relatively straightforward to fulfil
  this condition.
  Note that at the LHC, $\ptM$ is much larger than the $\ptjv$ values
  that will be of interest to us. 
  Therefore, while we take care to enforce
  Eq.~(\ref{eq:Sigmatilde-with-g1-g2}), the terms involving $\ptM$
  will in practice largely be irrelevant.
  Nevertheless the use of the modified logarithm is important also
  because it ensures that the $\Ltilde$ that enters in
  Eq.~(\ref{eq:Sigmatilde-with-g1-g2}) approaches zero for $\ptjv
  \gtrsim Q$.
  
\end{itemize}
Even with these conditions, there is some freedom in the matching
procedure.
This is closely related to the freedom we had in expressing the
efficiencies.
We will therefore consider three matching schemes, each of which is
the counterpart of one of our fixed-order efficiency prescriptions, so
as to facilitate the comparison with the NNLO results.

The first of our matching schemes is given by
\begin{multline}
 \Sigma_{\text{matched}}^{(a)}(\ptjv) = \\
   \left(\frac{\Sigmatilde_{\rm NLL}(\ptjv)}{\sigma_{0}}\right)^{Z}
    \Bigg[\sigma_0 + \Sigma_{1}(\ptjv) +\Sigma_2(\ptjv)
- Z \left(\Sigmatilde_{\rm NLL,1}(\ptjv)+\Sigmatilde_{\rm NLL,2}(\ptjv)\right)\\
      - 
      Z\frac{\Sigmatilde_{\rm NLL,1}(\ptjv)}{\sigma_0}
      \left(\Sigma_{1}(\ptjv) 
        - \frac{Z+1}{2} 
        \Sigmatilde_{\rm NLL,1}(\ptjv)\right)
 \Bigg]
   \>,  
\end{multline}
where $\Sigmatilde_{\rm NLL,i}(\ptjv)$ is the term of $\order{\as^i}$
relative to $\sigma_0$ in the fixed-order expansion of
$\Sigmatilde_{\rm NLL}$.
The factor $Z = \left(1-\frac{\ptjv}{\ptM}\right)$ is
necessary to satisfy Eq.~(\ref{eq:matching-at-kinematic-limit}), but
is largely irrelevant in practice. 
The corresponding jet-veto efficiency is
\begin{equation}
  \eff_{\text{matched}}^{(a)}(\ptjv) = 
  \frac{\Sigma_{\text{matched}}^{(a)}(\ptjv)}{\Sigma_{\text{matched}}^{(a)}(\ptM)}\,.
  \label{eq:efflogR}
\end{equation}
It is straightforward to verify that with this matching procedure
$\eff_{\text{matched}}^{(a)}(\ptjv)$ satisfies all three requirements listed at the
beginning of this Section.

The second matching scheme is identical except that we replace
$\Sigma_2(\ptjv)$ with $\bar \Sigma_2(\ptjv)$, in direct analogy with
scheme $b$ used for the fixed order efficiency definition,
\begin{multline}
 \Sigma_{\text{matched}}^{(b)}(\ptjv) = \\
   \left(\frac{\Sigmatilde_{\rm NLL}(\ptjv)}{\sigma_{0}}\right)^{Z}
    \Bigg[\sigma_0 + \Sigma_{1}(\ptjv) +{\bar \Sigma}_2(\ptjv)
- Z \left(\Sigmatilde_{\rm NLL,1}(\ptjv)+\Sigmatilde_{\rm NLL,2}(\ptjv)\right)\\
      - 
      Z\frac{\Sigmatilde_{\rm NLL,1}(\ptjv)}{\sigma_0}
      \left(\Sigma_{1}(\ptjv) 
        - \frac{Z+1}{2} 
        \Sigmatilde_{\rm NLL,1}(\ptjv)\right)
 \Bigg]
   \>,  
\end{multline}
with
\begin{equation}
  \eff_{\text{matched}}^{(b)}(\ptjv) = 
  \frac{\Sigma_{\text{matched}}^{(b)}(\ptjv)}{\Sigma_{\text{matched}}^{(b)}(\ptM)}\,.
  \label{eq:efflogR-matched}
\end{equation}

The third scheme that we consider is formulated directly in terms of
the jet-veto efficiency and is thus close in spirit to fixed-order
scheme $c$:
\begin{multline}
  \epsilon_{\text{matched}}^{(c)}(\ptjv) = \\
   \left( \epsilon_{\rm NLL}(\ptjv)\right)^{Z}
    \Bigg[1 + \epsilon_{1}(\ptjv) +\epsilon_2(\ptjv)
- Z \left( \epsilon_{\rm NLL,1}(\ptjv)+ \epsilon_{\rm NLL,2}(\ptjv)\right)\\
      - 
      Z \epsilon_{\rm NLL,1}(\ptjv)
      \left(\epsilon_{1}(\ptjv) 
        - \frac{Z+1}{2} 
         \epsilon_{\rm NLL,1}(\ptjv)\right)
 \Bigg]
   \>,  
\end{multline}
where, in accord with earlier definitions, we have
\begin{subequations}
  \begin{align}
    { \epsilon}_{\rm NLL}(\ptjv) &\equiv \frac{{\Sigmatilde}_{\rm NLL}(\ptjv)}{\sigma_{0}},\\
    { \epsilon}_{{\rm NLL},i}(\ptjv) &\equiv \frac{{\tilde
        \Sigma}_{{\rm NLL},i}(\ptjv)}{\sigma_{0}},\\
    \epsilon_{1}(\ptjv) &\equiv \frac{\bar \Sigma_{1}(\ptjv)}{\sigma_{0}},\\
    \epsilon_{2}(\ptjv) &\equiv \frac{\bar \Sigma_{2}(\ptjv)}{\sigma_{0}} -
    \frac{\sigma_{1}\bar \Sigma_{1}(\ptjv)}{\sigma_0^2}.
  \end{align}
\end{subequations}

The three schemes differ in two respects: firstly they treat
non-asymptotic terms (i.e.\ $\ptjv \sim Q$) differently; secondly,
they lead to different sets of subleading logarithms.
In particular, schemes $a$ and $b$ differ at the level of N$^4$LL
terms, while scheme $c$ differs from both of them by N$^3$LL terms.
We recall that NNLL differences will be probed by scale variations.

The above three schemes all belong to the family of multiplicative
matchings~\cite{DISresum}. Other schemes exist. In particular we have
also examined the log-$R$ scheme of Ref.~\cite{CTTW} and found that its
results are contained within the band defined by the three schemes
above.

\subsection{Matched results}

\begin{figure}[tbp]
\includegraphics[width=0.49\textwidth]{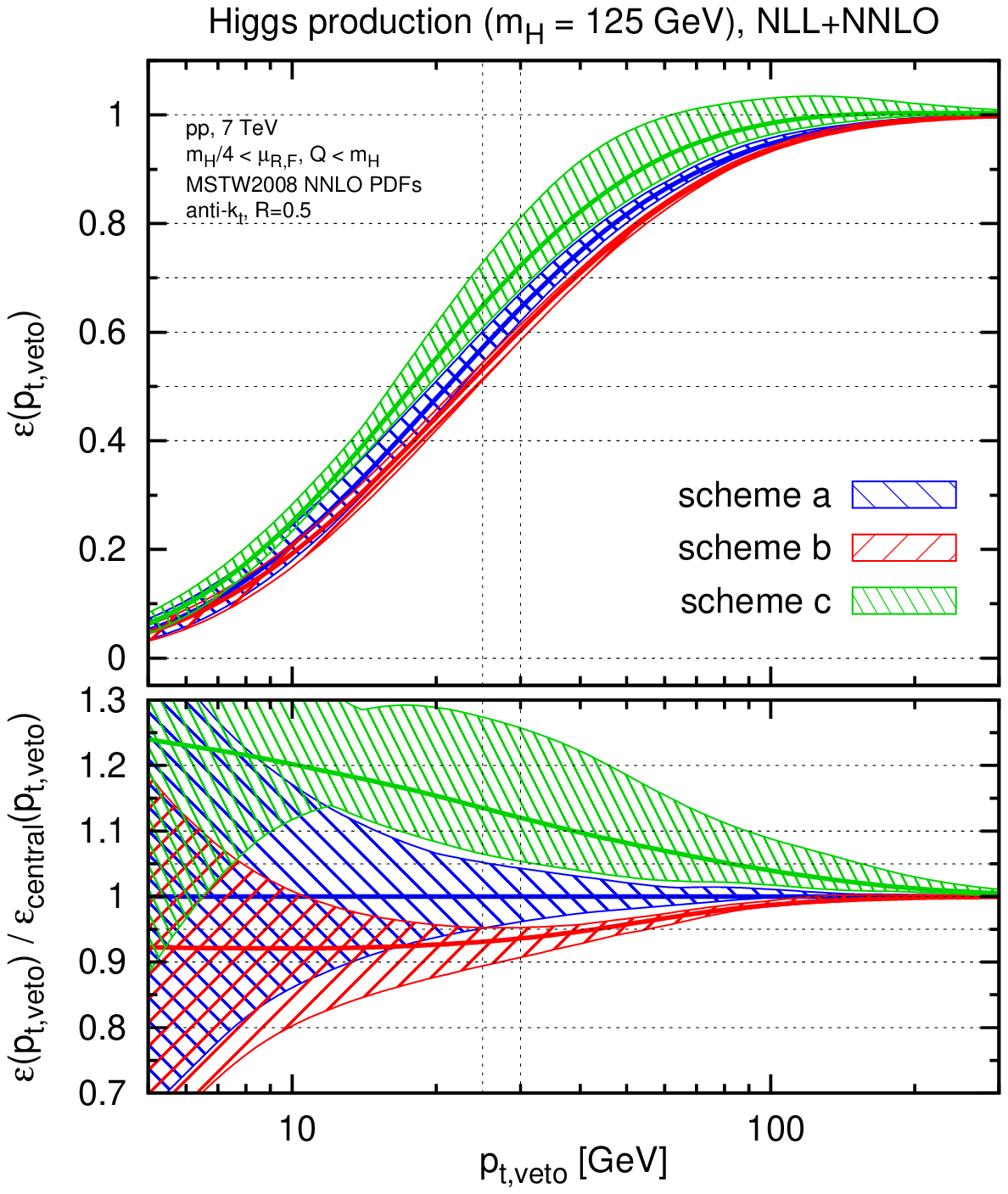}\hfill
\includegraphics[width=0.49\textwidth]{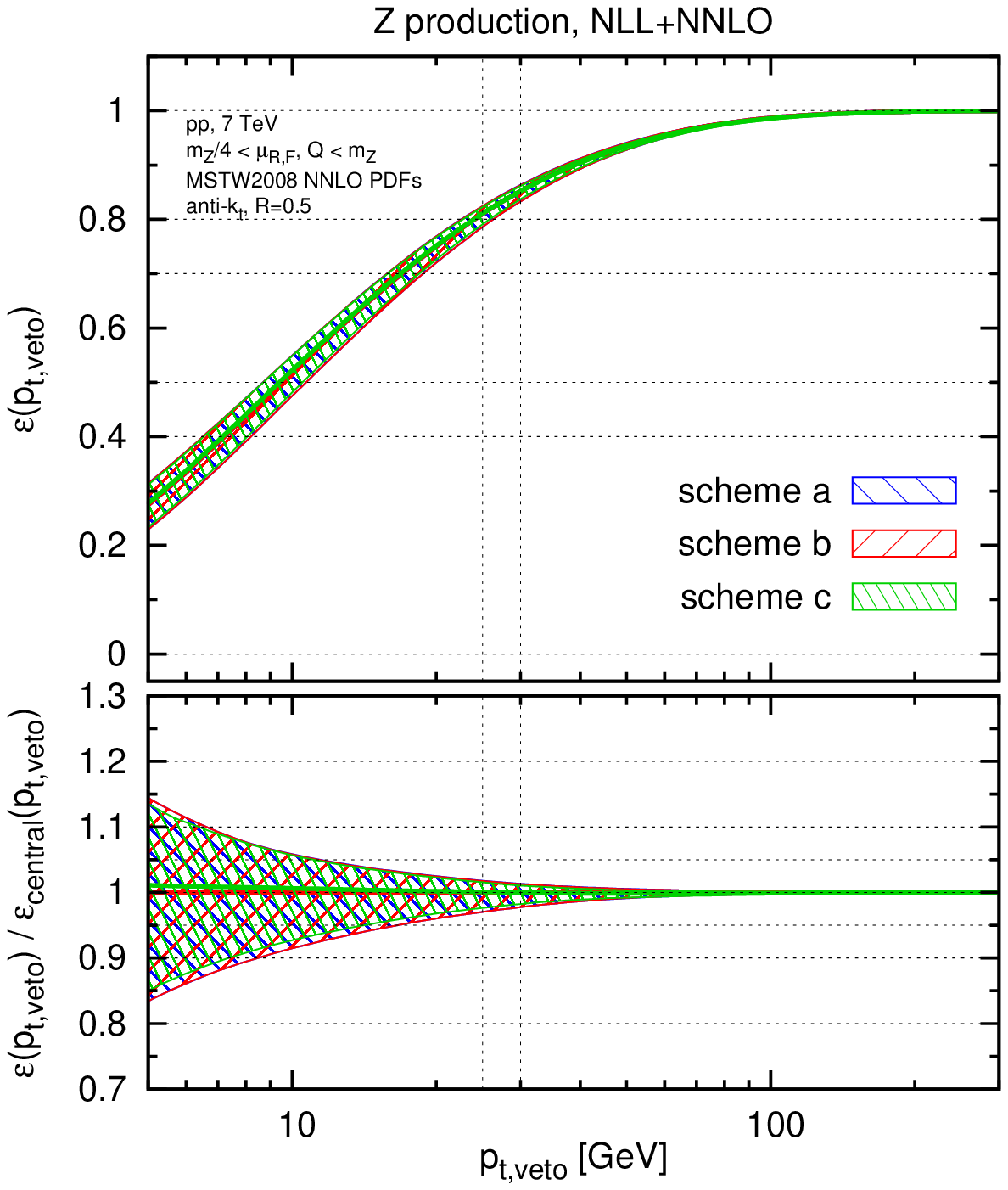}
\caption{NLL+NNLO jet-veto efficiency for Higgs (left) and $Z$-boson production
  (right) using three different matching prescriptions. 
  For each one, the thick solid line corresponds to the result obtained with $\mu_R =
  \mu_F = Q = M_{H/Z}/2$, while the
  band shows the scale uncertainty as obtained from the envelope of the
  choices of Eq.~(\ref{eq:RF-scale-choices}) and from $Q$-scale variation
  (taking $Q = \{M_B/4, M_B/2,M_B\}$ for  $\mu_R =  \mu_F =  M_{H/Z}/2$).
  The lower panels show the results normalised to the central scale
  choice for scheme $a$.
}
\label{fig:ptjv3matched}
\end{figure}

Figure~\ref{fig:ptjv3matched} shows the NLL+NNLO matched results for
each of the three schemes discussed above. 
Each band corresponds to the envelope of renormalisation, factorisation
and $Q$ scale variation.
Comparing to the pure resummed and fixed-order results of
figures~\ref{fig:pure-resummation} and \ref{fig:ptjv3NNLO}, one
observes that for larger $\ptjv$, the results display the features of
the fixed-order results, with similar uncertainties and, for scheme
$c$, the same artifact of $\eff > 1$ for certain scale choices.
As in the NNLO case, for Higgs production the uncertainty bands from
the three schemes mostly do not overlap, which serves to highlight the
importance of the use of different schemes for probing uncertainties.
For low $\ptjv$ values, the results coincide with the pure resummation,
though with uncertainty bands that are a little narrower, and that 
overlap even in the Higgs case.
In the region of intermediate $\ptjv \sim 25-30\GeV$, the central
value from the matched calculation is closer to that of the
fixed-order results than to the resummed results, but with a slightly
reduced uncertainty, indicating that resummation is at the edge of its
validity in this region.
Here too the bands from the different matching schemes fail to
overlap in the Higgs case.
Since the bands differ at most by NNNLL terms, this has implications
for the degree of improvement that one might expect when extending the
resummation from NLL to NNLL accuracy.

\begin{figure}[tbp]
\includegraphics[width=0.49\textwidth]{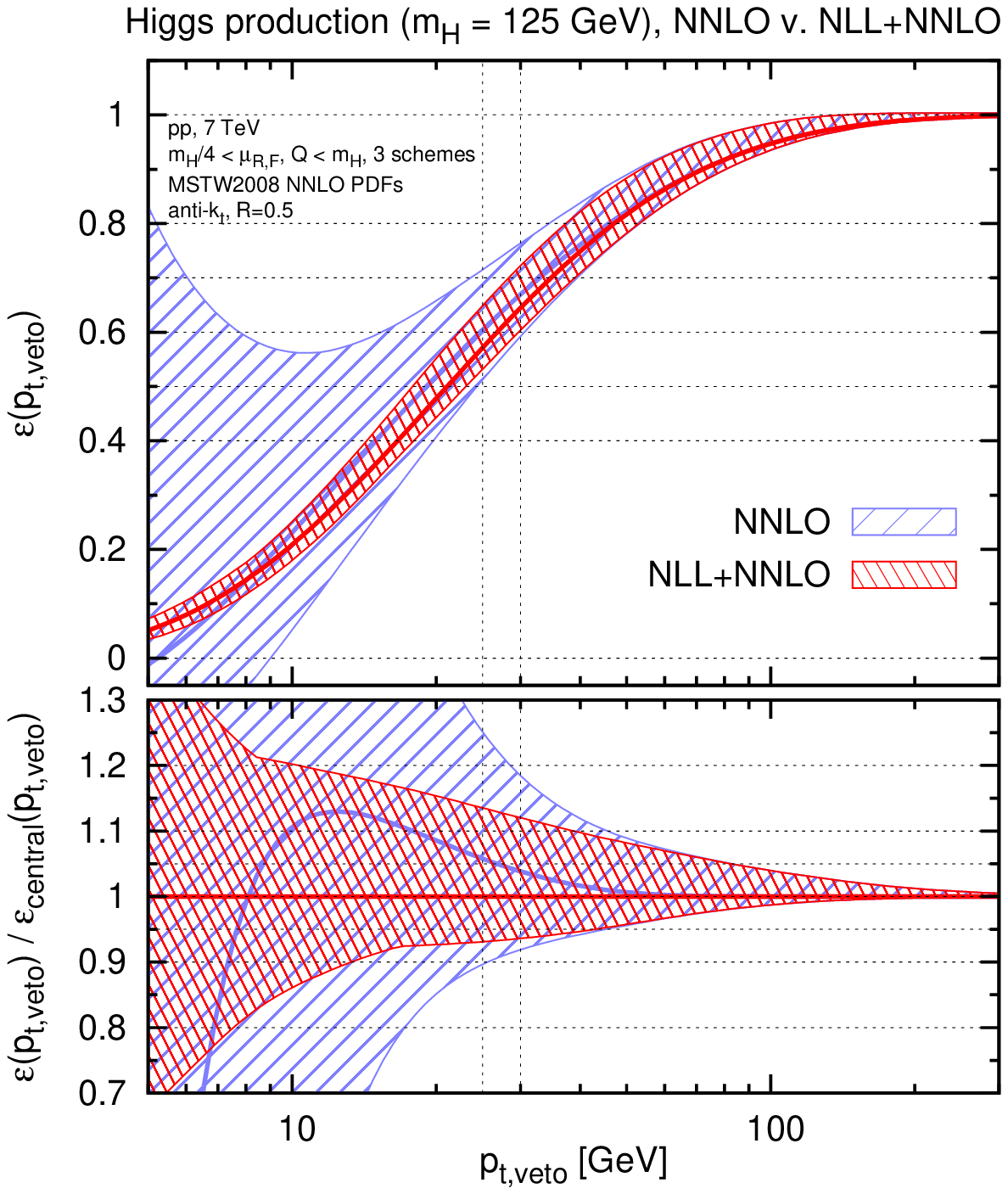}\hfill
\includegraphics[width=0.49\textwidth]{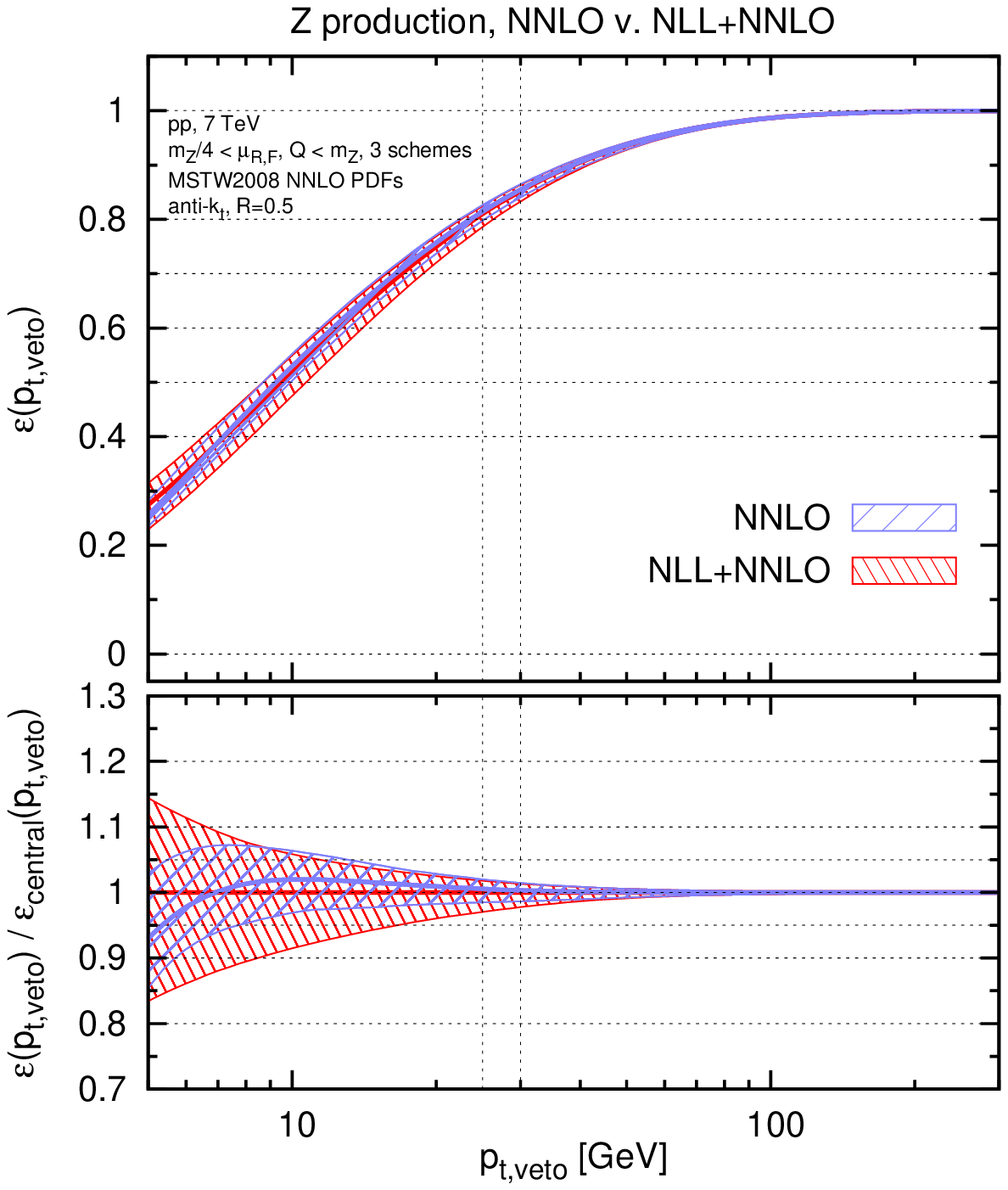}
\caption{
  Comparison of fixed-order (NNLO) and matched resummed
  (NLL+NNLO) predictions for the jet veto efficiencies in Higgs (left)
  and $Z$ production (right). 
  The uncertainties are those derived from the envelope method: for
  both fixed order and matched results they include renormalisation
  and factorisation scale
  uncertainties, as well as the scheme for defining the efficiency (or
  matching prescription). 
  In the matched case, there is additionally the uncertainty from the
  variation of $Q$.
  The lower panels show the ratio of the results to the central
  matched prediction.
}
\label{fig:ptjv-NNLO-v-matched}
\end{figure}

A direct comparison of the fixed-order and matched predictions
is to be found in Fig.~\ref{fig:ptjv-NNLO-v-matched}. Here the
uncertainty envelopes encompass the full scheme $a$ band as well as the
central values of the two other schemes.
This follows the procedure outlined in
section~\ref{sec:uncertainty-prescription} and it provides the
uncertainties that we shall use throughout the rest of the article.
The efficiencies for the two jet-veto thresholds used by ATLAS and
CMS, $25$ and $30\GeV$ respectively, are summarised in
table~\ref{tab:efficiencies}.
For Higgs production, one observes that  the absolute efficiencies are
about $3\%$ lower in 
the matched calculation as compared to the NNLO result (equivalent to
a relative $5\%$ reduction in the efficiency). The uncertainties are
somewhat more asymmetric in the matched calculation and in particular
the uncertainty towards lower efficiencies is reduced by about a
factor of two. For $Z$-boson production, the
uncertainties with matching are the same or larger as those of the
pure NNLO result. 
This surprising result may be because the resummation explicitly
involves the running of the coupling and thus, for low $\ptjv$,
directly probes the uncertainties associated with a perturbative
expansion whose coupling constant is somewhat larger than the
$\as(M_Z/2)$ that appears in the NNLO calculation. 

\begin{table}
  \centering
  \begin{tabular}{ccc}
    \multicolumn{3}{c}{Higgs production ($M_H = 125\GeV$)}\\[3pt]
    \toprule
    & NNLO & NLL+NNLO\\\midrule
     $\ptjv = 25\GeV$ & $60^{+11}_{-9}\%$ & $57^{+8}_{-4}\%$\\[4pt]
     $\ptjv = 30\GeV$ & $67^{+9}_{-8}\%$ & $64^{+8}_{-4}\%$\\
     \bottomrule
  \end{tabular}
  \hfill
  \begin{tabular}{ccc}
    \multicolumn{3}{c}{Z production}\\[3pt]
    \toprule
    & NNLO & NLL+NNLO\\\midrule
     $\ptjv = 25\GeV$ & $81^{+1}_{-2}\%$ & $81^{+1}_{-2}\%$\\[4pt]
     $\ptjv = 30\GeV$ & $85^{+1}_{-1}\%$ & $85^{+1}_{-2}\%$\\
     \bottomrule
  \end{tabular}

  \caption{Jet veto efficiencies and their uncertainties at NNLO and
    NLL+NNLO, for the values of $\ptjv$ used by ATLAS and CMS, shown
    for the anti-$k_t$ algorithm with $R=0.5$, and based on
    MSTW2008 NNLO PDFs. }
  \label{tab:efficiencies}
\end{table}

\section{Comparisons to other calculations}
\label{sec:compMC}
In this section we will complement our resummed matched study so far
with information from event generators and analytical boson-$p_t$
resummations. 
For brevity we concentrate on the case of Higgs production, using $M_H
= 125 \GeV$ throughout.
%

\subsection{Effects beyond the scope of matched calculations}

The matched calculation that we have performed applies to partons and
assumes infinite detector acceptance.
Experiments, however, measure hadrons, including the underlying event,
and have limited acceptance, notably for the rapidity of the jets.

\begin{figure}[tp]
\includegraphics[width=0.49\textwidth]{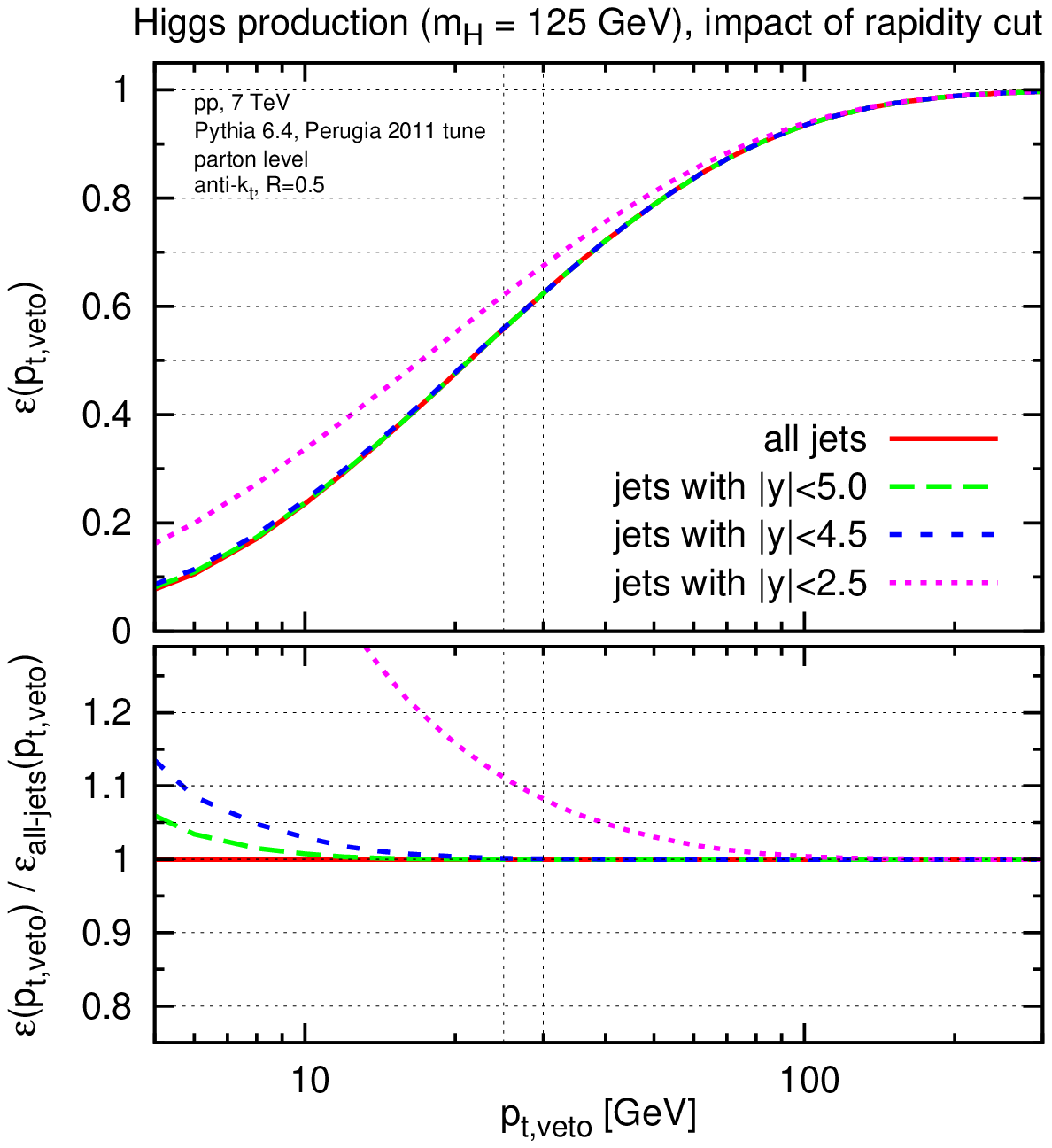}\hfill
\includegraphics[width=0.49\textwidth]{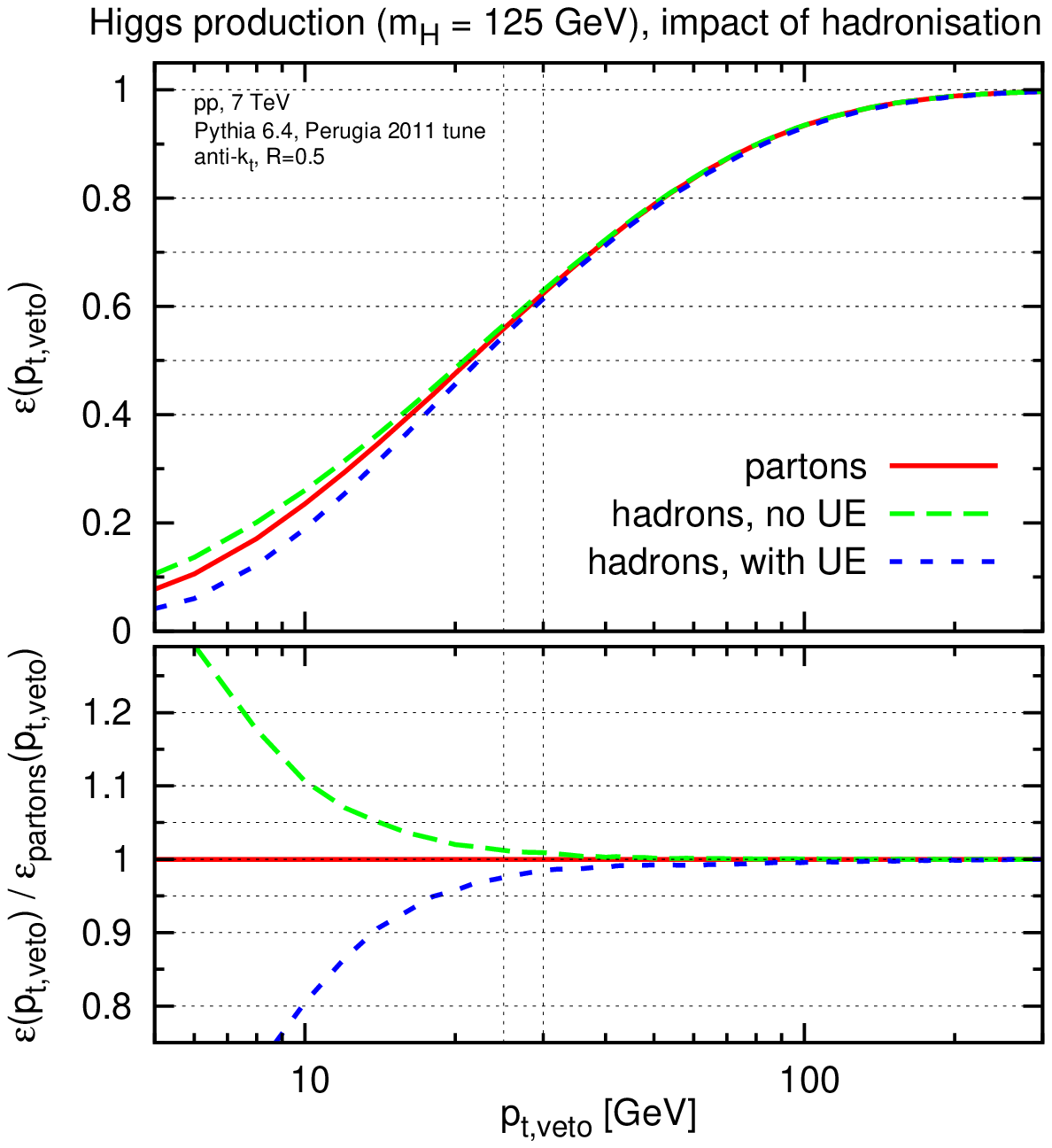}
\caption{
  Left: impact of a finite rapidity acceptance for jets on the
  jet-veto efficiency, as calculated with \Pythia 6.425.
  Right: impact of hadronisation and underlying event on the jet-veto
  efficiency.
  See text for further details.
}
\label{fig:MC-rapdep+levels}
\end{figure} 

To investigate these two effects we have taken events generated by
\Pythia~6.4~\cite{Sjostrand:2006za} with the Perugia~2011
tune~\cite{Skands:2010ak}. Jet clustering for the results in this
section is performed with FastJet~\cite{FastJet}.
Fig.~\ref{fig:MC-rapdep+levels} (left) shows the impact of considering
jets only within some finite rapidity acceptance.
One sees that for the choices used by ATLAS and CMS, $|y|<4.5$ and
$y<|5.0|$ respectively, the veto efficiencies are almost identical to
those with full acceptance in the practically relevant range of $\ptjv$.
We have confirmed that this pattern holds also in fixed-order
calculations.
In contrast, if one applies a jet veto only in a more restricted
rapidity region, e.g.\ $|y|< 2.5$, there are substantial differences.

The right-hand plots of Fig.~\ref{fig:MC-rapdep+levels} show the
impact of non-perturbative effects.
While not entirely negligible, both hadronisation and the underlying
event have an impact that is somewhat smaller than the uncertainties
on the matched calculation, at least for the range of $\ptjv$ values
of practical interest.
This contrasts with the situation for variables that receive
contributions from all hadrons in the event, such as the beam thrust
or transverse energy flow, which see large contributions from the
underlying event (cf.\ the studies in
Ref.~\cite{Papaefstathiou:2010bw}).

\subsection{Comparisons with \POWHEG and \HqT}

Both ATLAS~\cite{ATLAS:2011aa} and CMS~\cite{Chatrchyan:2012ty} make
use of \POWHEG~\cite{Alioli:2008tz} for estimating their jet-veto
efficiencies, interfaced to \Pythia 6.4 (this is stated explicitly by
ATLAS, we assume it to be the case for CMS).
The ATLAS collaboration additionally reweights the events so as to
ensure that the Higgs boson $q_t$ spectrum coincides with that of the
\HqT program~\cite{Bozzi:2005wk}.
In this section we compare our results to these various tools.

\begin{figure}[pt]
  \centering
  \includegraphics[width=0.49\textwidth]{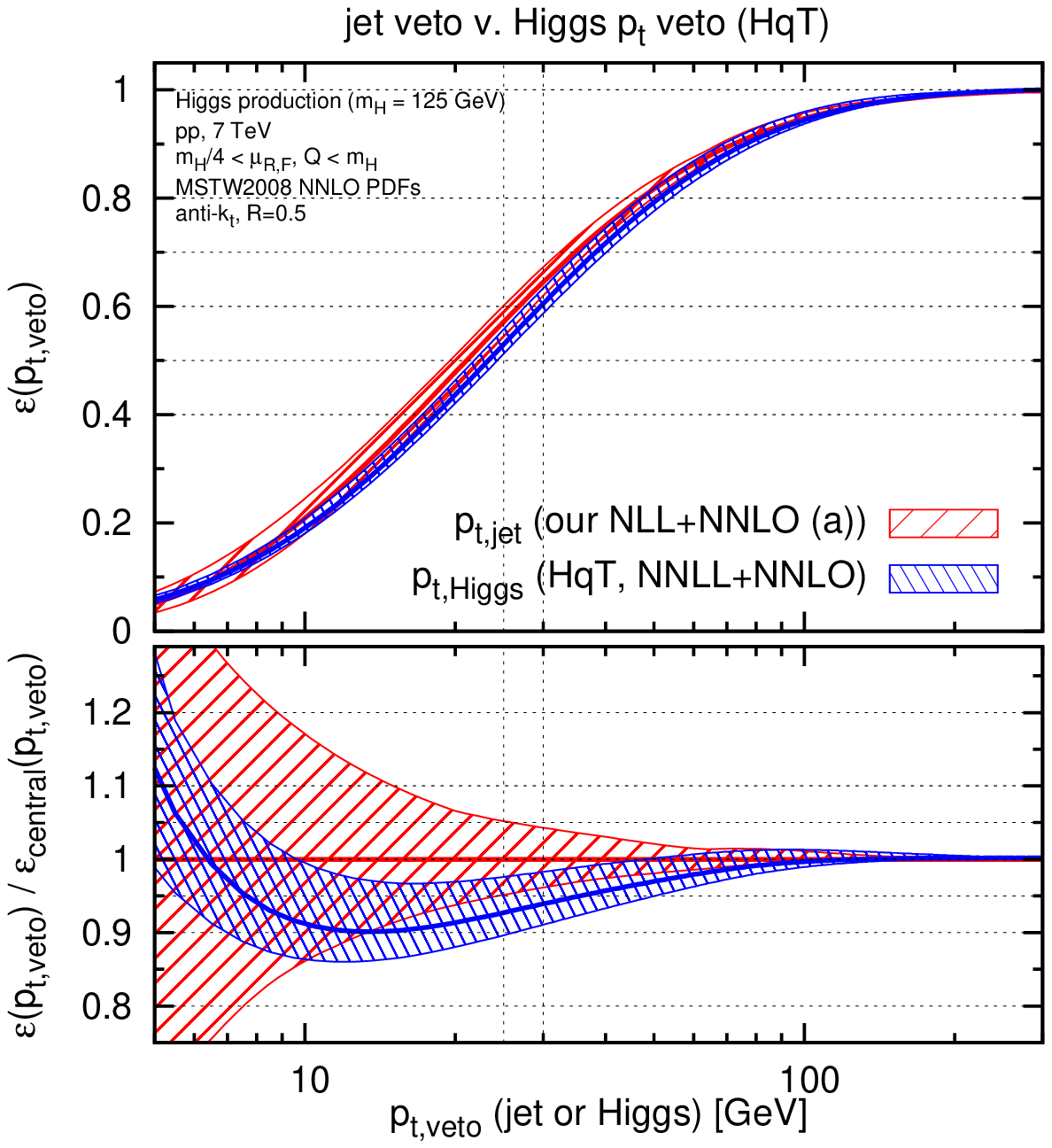}\hfill
  \includegraphics[width=0.49\textwidth]{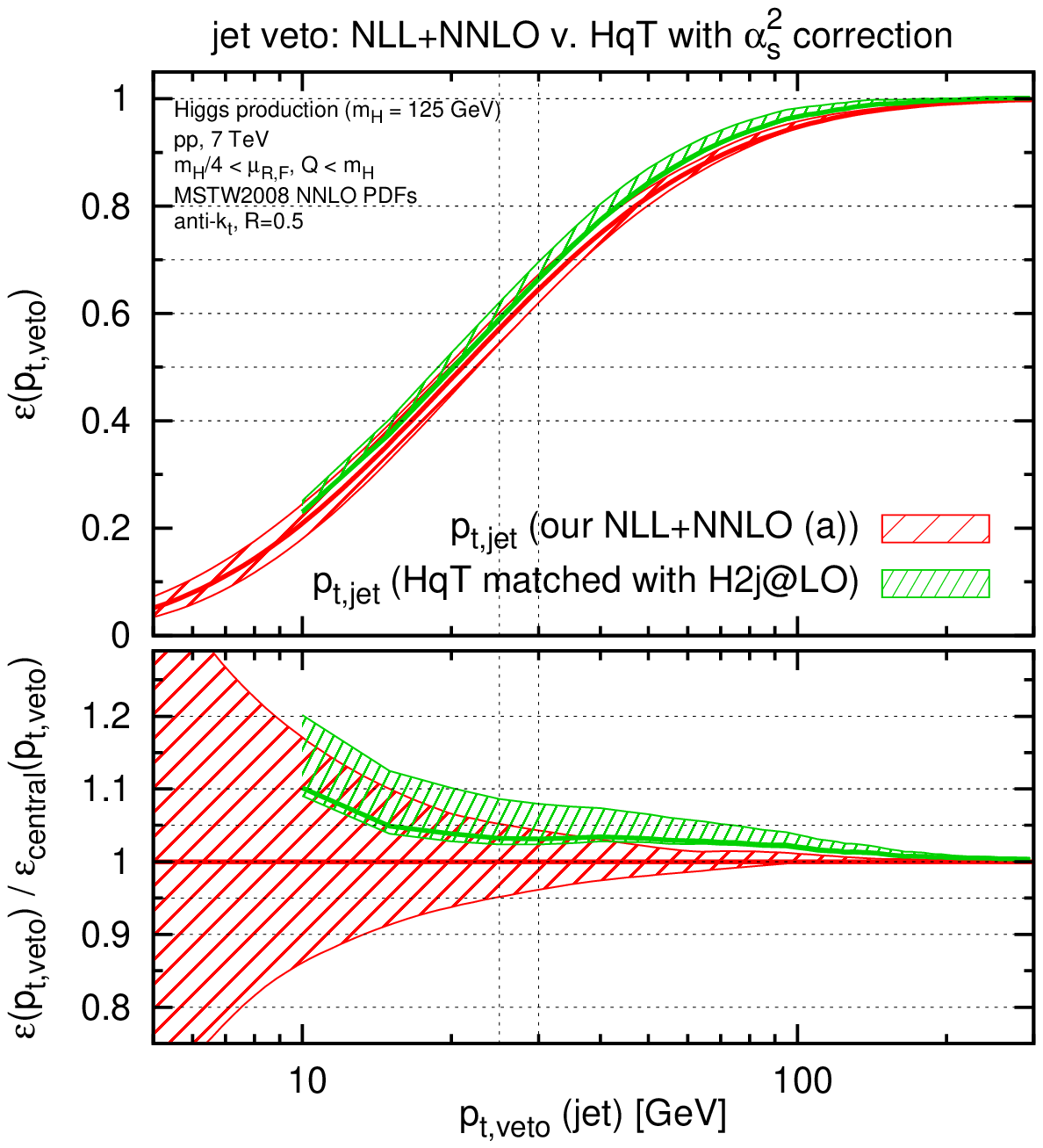}
  \caption{Left: comparison of our result for the jet veto efficiency with the
    calculation of a 
    Higgs-boson transverse momentum veto efficiency obtained with \HqT.
    For the purpose of the comparison, since \HqT provides only a
    single matching scheme, we restrict the uncertainty band on our
    results to use just matching scheme ($a$).  
    Right: comparison of our jet veto result with the jet veto result
    obtained by 
    correcting \HqT with the (relative) order $\as^2$ Higgs+2-jet 
    contribution, as in Eq.~(\ref{eq:H2j-LOmatching}). Again our band
    includes just scheme ($a$). }
  \label{fig:HqT-v-us}
\end{figure}

We start, figure~\ref{fig:HqT-v-us} (left), by comparing the \HqT NNLL+NNLO
result for a veto on the Higgs-boson transverse momentum to our NLL+NNLO
result for the jet veto.
Part of the purpose of this comparison is to examine the relative sizes
of the uncertainties and the benefit to be had from a NNLL
resummation.
In order for the comparison to be consistent with \HqT, which has just
a single matching scheme, our uncertainty band here includes only
scale variation with scheme (a), but not the central values from the
other schemes.
At low values of $\ptjv \lesssim 20 \GeV$, it appears that there are
clear benefits to be had from the NNLL resummation, with a significant
reduction in the uncertainty as compared to a NLL result.
On the other hand, for the practically relevant region, $\ptjv \sim
30\GeV$, the bands are quite similar in size.
This is perhaps not surprising since at this scale resummation only
just starts to become relevant.

The modest size of the differences between a transverse momentum veto
on the Higgs-boson and one on the jet is to be expected given that
the two observables differ only from $\order{\as^2}$ onwards.
An interesting cross-check of the consistency between the two
approaches can be obtained by taking the \HqT calculation of the Higgs
transverse-momentum and correcting it with the $\as^2$ difference
between Higgs and jet transverse momentum distributions.
Some care is needed in order not to introduce divergences at low
$p_t$, and we use the following matching procedure
\begin{equation}
  \label{eq:H2j-LOmatching}
  \eff_{\jet}^{\text{(HqT+H2j@LO)}}(\ptjv) =
  \left( 1 +
    \frac{\Sigma_{2,\jet}(\ptjv) - \Sigma_{2,\Higgs}(\ptjv)}{\sigma_0}
  \right)
  \eff_{\Higgs}^{\text{(HqT)}}(\ptjv)\,,
\end{equation}
involving the difference of $\Sigma_2$ terms for the jet-$p_t$ and
Higgs-$p_t$ calculations  normalised to $\sigma_0$, as well as the Higgs
$p_t$-veto efficiency, $\eff_{\Higgs}^{\text{(HqT)}}(\ptjv)$.
While such a matching does break the resummation accuracy, one may
still expect it to have some meaning in the region of intermediate
transverse momenta.
The result is shown in Fig.~\ref{fig:HqT-v-us} (right) as the band labelled
``\HqT matched with H2j@LO'', using the MCFM calculation of H+2-jets
at LO. It is in reasonable agreement with our NLL+NNLO
calculation. This helps illustrate the consistency between the
different tools.
The study can be extended to NLO in the H+2-jet
calculation~\cite{Campbell:2010cz}, and the results appear broadly
similar, though care is needed in the treatment of technical infrared
cutoffs in the H+2-jet calculation.

\begin{figure}[tp]
\includegraphics[width=0.49\textwidth]{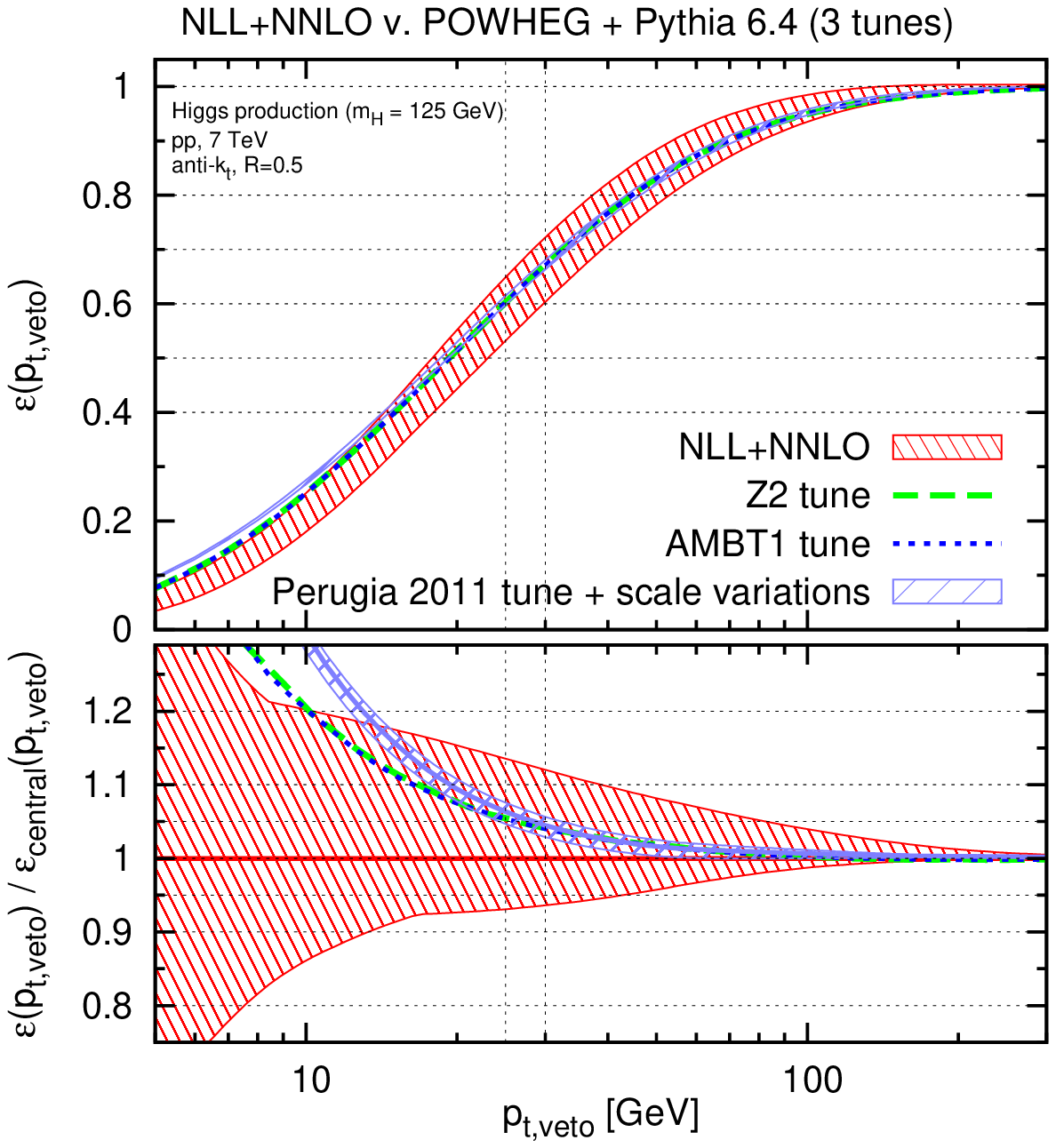}\hfill
\includegraphics[width=0.49\textwidth]{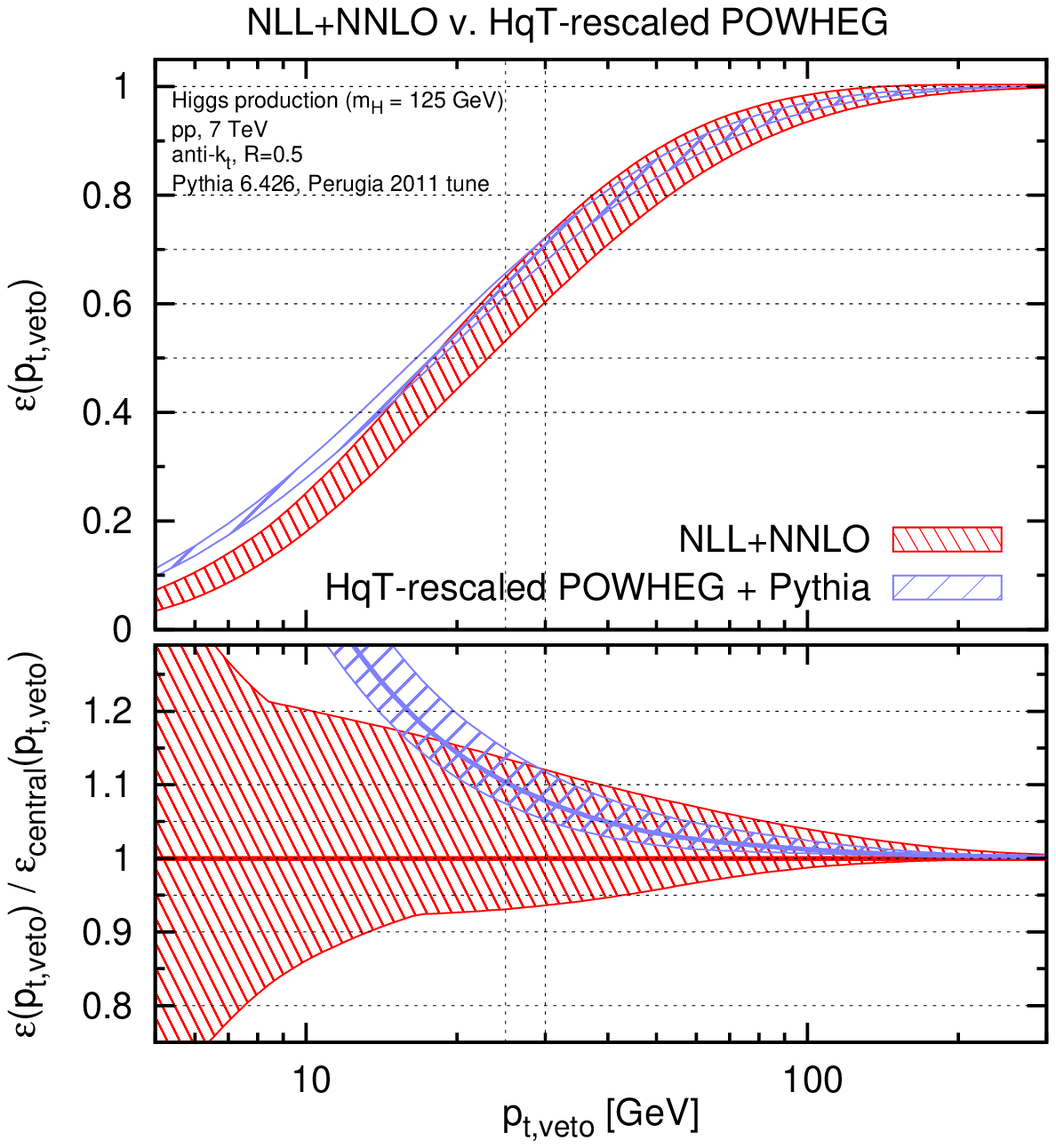}
\caption{ %
  Left: comparison of \POWHEG + \Pythia 6.426 with our NLL+NNLO
  predictions; %
  results are shown for three Pythia tunes. One of them, Perugia 2011,
  is displayed as a band, formed by the envelope of renormalisation
  and factorisation scale variations
  (Eq.~(\ref{eq:RF-scale-choices})). 
  Right: comparison of the \HqT-rescaled \POWHEG + \Pythia combination
  (Perugia 2011 tune) with our NLL+NNLO results.
  \POWHEG and \HqT are always used with MSTW2008NNLO PDFs. }
\label{fig:POWHEG}
\end{figure}

Let us now turn to \POWHEG.\footnote{We used revision 1683.} Here we
use default renormalisation and 
factorisation scales equal to $M_H/2$, as in our fixed order and
resummed calculations.
Fig.~\ref{fig:POWHEG} (left) shows the results, showered with 3
different commonly used tunes of \Pythia, Perugia~2011,
Z2~\cite{Field:2010bc} and AMBT1~\cite{ATLAS:2010ir}.
For the $p_t$ values of interest, all three are contained within the NLL+NNLO
uncertainty band, though systematically above its central value. 

To probe the uncertainties in \POWHEG plus
\Pythia, we followed the suggestion in
refs.~\cite{Dittmaier:2012vm,Campbell:2012am}, and varied
renormalisation and factorisation scale independently around $M_H/2$
as described in Eq.\eqref{eq:RF-scale-choices} and fixed the parameter
${\tt hfact}$ to $h=M_H/1.2$.
With these choices the Higgs transverse momentum spectrum of \PWGPYT
yields a reasonable agreement in shape with that of \HqT.
We recall that in \POWHEG it is possible to split the real radiation
contribution $R$ into a shower part $R^s$ and a finite one $R^f$, with the
requirement that $R^s\to R$ at small transverse momenta. 
Choosing a finite ${\tt hfact}$ (rather than ${\tt hfact}=\infty$)
means that $R^s = h^2/(h^2+p_{T,H}^2)R$ and $R^f = p_{T,H}^2/(h^2+p_{T,H}^2)
R$. Therefore this choice mainly affects the large transverse momentum
region. Since the transverse momentum spectrum of shower events is not   
affected by scale variations, while that of finite events is, this
splitting provides one way of assessing of the uncertainties in
\POWHEG.

In Fig.~\ref{fig:POWHEG} (left) this is illustrated as a band for the
case where \POWHEG is interfaced to \Pythia with the Perugia 2011 tune.
This band is significantly narrower than our uncertainty band and one
also observes that for low values of $\ptjv$, it does not quite
encompass the different tunes. 
We tend to believe that the \POWHEG uncertainty band represents an
underestimate of the true uncertainties. 
This suggests that it might be valuable to investigate other sources
of uncertainty, for example variations of {\tt hfact}.

The right-hand plot of Fig.~\ref{fig:POWHEG} shows the impact of
reweighting events so as to ensure that the final Higgs $p_t$
distribution agrees with the output of the NNLL+NNLO \HqT program.
In this procedure, every (originally unweighted) event $i$ from
\PWGPYT is assigned a weight $w_i$
\begin{equation}
  \label{eq:reweight}
  w_i = \left. 
    \left(
    \left. \frac{d\sigma^{\text{HqT}}}{dp_{t,H}}\right|_{p_{t,H}^{(i)}}\right)
  \right/
    \left(
    \left. \frac{d\sigma^{\text{POWHEG+Pythia}}}{dp_{t,H}}\right|_{p_{t,H}^{(i)}}
    \right)
\end{equation}
where $p_{t,H}^{(i)}$ is the transverse momentum of the Higgs boson in
event $i$.
The result of the weighting for the main tune we consider, Perugia
2011, is to increase the efficiency a little, bringing it close the
upper edge of our (full) band.
The reweighted \PWGPYT band is broader than that before reweighting,
but still smaller than ours.
This is natural given that \HqT considers just a single matching
scheme as compared to our multiple schemes.
Note also that the reweighting of \PWGPYT with \HqT breaks the
NNLO accuracy of the \HqT calculation, unlike the procedure of
Eq.~(\ref{eq:H2j-LOmatching}).

Finally, we remark that while here we have considered just the \PWGPYT
results for NLO-matched parton showers, comparisons with
\MCatNLO~\cite{Frixione:2002ik} showered with
\textsc{herwig}~\cite{Corcella:2002jc} are to be found in
Ref.~\cite{Dittmaier:2012vm}.

\section{Conclusions}
\label{sec:conclu}

In this article we have examined resummations and matching to fixed
order calculations for jet vetoes in Higgs-Boson and $Z$-boson
production.
The NLL resummation has a particularly simple form, reducing to just a
Sudakov form factor.
Such a simplification had been observed in the past for the two-jet
rate with the Cambridge algorithm in $e^+e^-$ collisions.
Here the result holds for a broad class of jet algorithms,
independently of the jet radius.

To make phenomenological use of our result it was necessary to match
the resummation with a fixed-order calculation, especially because
experimentally relevant jet-veto scales are at the edge of the region
where resummation starts to become relevant.
One may express the result either as a jet-vetoed cross section or a
jet-veto efficiency. 
We chose to concentrate on the latter for two reasons: firstly,
the efficiency is closely related to the concept behind a Sudakov
resummation; secondly, the perturbative structure of the efficiency
provides useful handles when estimating uncertainties.
The uncertainties on the efficiency should, we believe, be essentially
uncorrelated from those on the total cross section, making it
straightforward to also derive uncertainties on vetoed cross sections.

For $Z$-boson production, the jet-veto efficiency turns out to be
rather stable.
For Higgs production, however, even if the NLL+NNLO matching provides
some degree of stabilisation of the results relative to just NNLO, the
final uncertainties are not negligible, cf.\
table~\ref{tab:efficiencies}.
The problem can be traced back to ambiguities in how one formulates
the fixed-order result for the efficiency and is closely connected
with the poor convergence of the perturbative series for the total
cross section. 
Essentially, considering that the known NNLO corrections to the total
Higgs cross section are about $20\%$, one can expect that the unknown
NNLO corrections to the 1-jet rate can also easily be of order $20\%$
(and distinct from the uncertainty on the total cross section, associated with
unknown NNNLO terms), which then inevitably propagates into the
jet-veto efficiency.
This issue is generally not addressed in existing calculations, with
the exception of Ref.~\cite{Stewart:2011cf}, which approaches it from
a somewhat different point of view.
From the perspective of resummations, such effects are N$^3$LL, and so
appear to be beyond current technology.
Steps towards NNLL accuracy, specifically terms related to the choice
of jet definition, are discussed in the appendix.

Given the NLL+NNLO calculation, there were various phenomenological
questions that could be addressed.
On one hand, it was useful to establish how directly the result could be
applied to data.
Effects of potential relevance include finite detector acceptance and
non-perturbative corrections.
Based on Monte Carlo simulations, both appear to be modest compared to
other remaining uncertainties.
We also compared our results to the predictions from tools used by the
LHC experiments, such as the NNLL+NNLO Higgs-boson $p_t$ resummations
from the \HqT program and the NLO parton-shower generator \POWHEG
(with \Pythia), with or without rescaling corrections from the \HqT
program.
We observe a pattern of slightly higher efficiencies with these tools
than in our NLL+NNLO calculation, but all remain consistent to within
uncertainties.

\section*{Acknowledgements}
We thank Simone Alioli, Babis Anastasiou, Emanuele Di Marco, G\"unther
Dissertori, Massimiliano Grazzini, Pier Fracesco Monni, Paolo Nason,
Vivek Sharma and Maiko 
Takahashi for fruitful discussions and Matthias Neubert and Thomas
Becher for prompting us to complete this work in a timely fashion and
for comments on the manuscript.
We are grateful to CERN (AB and GZ), the GGI Institute (AB and GZ) and
ETH Z\"urich (GPS and GZ) for hospitality while part of this work was
carried out. This work was initiated while AB was at the ETH Z\"urich.
GZ is supported by the British Science and Technology Facilities
Council. 
The work of GPS is supported in part by the Agence Nationale de la
Recherche under contract ANR-09-BLAN-0060.
Both acknowledge the support also of grant PITN-GA-2010-264564 from the
European Commission.


\appendix

\section{NNLL $R$-dependent terms}
\label{sec:NNLL-R-dependence}

The \CAESAR paper~\cite{caesar} was primarily concerned with NLL
resummation. Nevertheless, it did provide certain formulae that are of
use in a NNLL context, as part of the verification of the particular
classes of terms that could be neglected in a NLL calculation.

It is our understanding that these formulae provide key pieces that
are missing in order to relate existing NNLL resummations for the
Higgs or vector boson $p_t$ distribution to a NNLL resummation of the
jet veto efficiency and specifically the dependence of the NNLL
corrections on the jet algorithm and radius.

While it is beyond the scope of this article to provide a full NNLL
analysis of the jet veto efficiency, we believe it is useful,
nevertheless, to make the NNLL $R$-dependent results available.

There are two sources of correction, related to the fact that the emission
of two soft-collinear gluons can be separated into two parts: an
independent-emission term and a correlated emission term, the latter
being non-zero only when the two gluons are close in rapidity:
\begin{equation}
  \label{eq:indep-correl}
  \cM^2_{gg}(k_1, k_2) = \cM_g^2(k_1)\, \cM_g^2(k_2)  +
  \widetilde \cM^2_{gg}(k_1, k_2)\,,
\end{equation}
where $\cM^2_{g}(k)$ is the factorised matrix element for the emission
of a single gluon and 
$\widetilde \cM^2_{gg}(k_1, k_2)$ is the correlated-emission
part of two-gluon emission matrix element in the limit where both are soft and
collinear with respect to the beam direction. 
There is also a corresponding correlated-emission component for the
production of a 
quark-antiquark pair, $\widetilde \cM^2_{q\bar q}$. We will use
$\widetilde \cM^2$ to denote $\widetilde \cM^2_{gg} + 2\widetilde
\cM^2_{q\bar q}$. 
For double soft (and collinear) emission off a quark line, the
non-trivial correlated part of the matrix element was given as early
as Refs.~\cite{Berends:1988zn, Dokshitzer:1992ip} and was rederived
for more general event structures in
Refs.~\cite{Campbell:1997hg,Catani:1999ss}.
Apart from colour factors ($C_F^2$, $C_F C_A$, $C_F T_R n_f$
respectively for independent and correlated gluon and quark-pair
emission from a quark line; $C_A^2$, $C_A^2$ and $C_A T_R n_f$ for
emission from a gluon line), the matrix elements are the same
regardless of whether the gluon pair or $q\bar q$ pair are emitted from a quark
line or a gluon line, as can be seen clearly in the formulae of
Ref.~\cite{Catani:1999ss}.

\subsection{Correlated emission}

We start with the correlated-emission component, and in particular
from Eq.~(D.8) of the \caesar paper, which provides the NNLL
contribution associated with the presence of any number of independent
emissions and one correlated pair:
\begin{multline}
  \label{eq:cF1correl}
  \cF^\mathrm{correl} = \exp\left(-\int_{\epsilon v}^v [dk]
    \cM^2_{g,rc}(k)\right) \times\\\times \sum_{n=0} \frac{1}{n!} \left(
    \prod_{i=1}^n
    \int_{\epsilon v} [dk_i]  \cM^2_{g,rc}(k_i)\right)
   \frac1{2!} \int [dk_a][dk_b] 
   {\widetilde \cM}_{gg,rc}^2(k_a,k_b) \times\\
  \times \left[\Theta(v - V(k_1,\ldots,k_n,k_a,k_b)) -
    \Theta(v - V(k_1,\ldots,k_n,k_a+k_b)) \right] \,.
\end{multline}
This formula can be understood as the correction that arises when an
observable is sensitive to the kinematics of the individual $a$ and
$b$ partons rather than just their sum.
For the resummation of a sufficiently inclusive quantity, notably the
boson $p_t$ distribution, it is explicitly zero.
Consequently, the calculation of this term should allow one to relate the
jet-veto resummation to the boson $p_t$ resummation.\footnote{We will
  not discuss here the effects of hard collinear emission off two
  different legs~\cite{Catani:2010pd}, which can introduce an extra
  source of differences between boson and jet $p_t$ resummations.}

For compactness we have not explicitly written the $\{\tilde p\}$
argument to $V(\ldots)$ in Eq.~(\ref{eq:cF1correl}).
We work in a limit where $\as \ll 1$, $\ln 1/v \gg 1$ and $\as
\ln 1/v$ is finite.
The parameter $\epsilon$, which serves as a regularisation cutoff, is
to be taken $\epsilon \ll 1$, but also such that $\as \ln 1/\epsilon
\ll 1$.
In this limit the phase space integrals $[dk_i]$ essentially extend up
to a rapidity $|y_i| \lesssim \ln 1/v$ and the dependence of the precise
upper rapidity limit on a given parton's transverse momentum will turn
out to be irrelevant at our accuracy.
Additionally, to within a factor $\order{\epsilon}$, all emissions
have a $p_t \sim v M_B \equiv \ptjv$ ($v = \ptjv/M_B$ and $M_B$, we
recall, is the boson mass).
The matrix elements include a subscript ``$rc$'' to indicate that the
strong coupling is to be evaluated at a scale of order $v M_B$.

A NNLL contribution arises only when all emissions $i = 1 \ldots n$
are well separated from each other and from $a$ and $b$.
For the case of the hardest jet's $p_t$, we then have 
\begin{subequations}
  \begin{align}
    \Theta(v - V(k_1,\ldots,k_n,k_a,k_b)) &=
    \left[\prod_{i=1}^n
    \Theta(v - V(k_i)) \right] \Theta(v - V(k_a,k_b)),\\
    \Theta(v - V(k_1,\ldots,k_n,k_a + k_b)) &= 
    \left[\prod_{i=1}^n \Theta(v - V(k_i)) \right]    
    \Theta(v - V(k_a + k_b))\,.
  \end{align}
\end{subequations}
Making use of the fact that 
\begin{equation}
  \label{eq:sum-over-emsns}
  \sum_{n=0} \frac{1}{n!} \left(
    \prod_{i=1}^n
    \int_{\epsilon v} [dk_i]  \cM^2_{g,rc}(k_i) \Theta(v - V(k_i))\right) 
  =
  \exp\left(\int_{\epsilon v}^v [dk] \cM^2_{g,rc}(k)\right) \,,
\end{equation}
we can then rewrite Eq.~(\ref{eq:cF1correl}) simply as
\begin{equation}
  \label{eq:cF1correl-just2}
  \cF^\mathrm{correl} = 
   \frac1{2!} \int [dk_a][dk_b] 
   {\widetilde \cM}_{gg,rc}^2(k_a,k_b) \left[\Theta(v - V(k_a,k_b)) -
    \Theta(v - V(k_a+k_b)) \right] \,,
\end{equation}
i.e.\ it reduces to a pure two-gluon result (with a running-coupling).

It is simplest to first evaluate the leading $R$-dependence of this
formula in the limit of a small jet radius $R$.
When the angle between the two partons $a$ and $b$ is
small, $\Delta_{ab}^2 = (y_a - y_b)^2 + (\phi_a - \phi_b)^2 \ll 1$, we
can write the phase-space and matrix element as
\begin{multline}
  \label{eq:PSME-small-angle}
  [dk_a][dk_b] {\widetilde \cM}_{gg,rc}^2(k_a,k_b) =
  \frac{2 C \as(k_{t,ab}) }{\pi} \frac{dk_{t,ab}}{k_{t,ab}} \frac{d\phi_{ab}}{2\pi} dy_{ab} 
  \times \\ \times
  \frac{\as(k_{t,ab} \Delta_{ab})}{\pi} \frac{d\Delta_{ab}}{\Delta_{ab}}
    dz \left( P_{gg}(z) + 2 n_f P_{qg}(z) \right)\,,
\end{multline}
where $C = C_F$ or $C_A$ depending on the nature of the beam partons,
$k_{ab} \equiv k_a + k_b$ and the $P_{gg}$ and $P_{qg}$ are the
real parts of the usual leading order splitting functions:
\begin{subequations}
  \begin{align}
    P_{gg}(z) &= 2C_A \left(\frac{z}{1-z} + \frac{1-z}{z} + z(1-z)\right)\,,\\
    P_{qg}(z) &= T_R\left(z^2 + (1-z)^2\right)\,.
  \end{align}
\end{subequations}
With these variables, for $\Delta_{ab} \ll 1$, the difference of $\Theta$ functions in
Eq.~(\ref{eq:cF1correl-just2}) reduces to
\begin{multline}
  \label{eq:thetas-simplify}
  \left[\Theta(v - V(k_a,k_b)) - \Theta(v - V(k_a+k_b)) \right] = 
  \\  \Theta(\Delta_{ab} - R) \left[ \Theta(v M_B - \max(z,1-z)k_{t,ab})
    - \Theta(v M_B - k_{t,ab})\right]\,,
\end{multline}
as long as one restricts one's attention to jet algorithms from the
generalised-$k_t$ family.
We can now rewrite Eq.~(\ref{eq:cF1correl-just2}) as 
\begin{multline}
  \label{eq:cF1correl-smallR-simple}
  \cF^\mathrm{correl} = 
  \frac1{2!} 
  \int
    \left(\frac{2 C \as(k_{t,ab}) }{\pi}  \frac{d\phi_{ab}}{2\pi} dy_{ab} \right)
  \left(\frac{d\Delta_{ab}}{\Delta_{ab}} \Theta(\Delta_{ab} - R)\right)
  \times \\ \times
  \frac{\as(k_{t,ab} \Delta_{ab})}{\pi}
    dz \left( P_{gg}(z) + 2 n_f P_{qg}(z) \right) 
    \times \\ \times
    \left(\frac{dk_{t,ab}}{k_{t,ab}}
     \left[ \Theta(v M_B - \max(z,1-z)k_{t,ab})
      - \Theta(v M_B - k_{t,ab})\right]\right)\,.
\end{multline}
A first point is that $k_{t,ab}$ will be limited to be of order $v
M_B = \ptjv$. 
Secondly, while we take $R < \Delta_{ab}  \ll 1$, we will still assume that
$\as \ln R$ is negligible.
Accordingly we can replace each of the running couplings with $\as(\ptjv)$. 
This puts us in a position to carry out the integrations in each of
the three lines of Eq.~(\ref{eq:cF1correl-smallR-simple})
independently.
The contents of the first round brackets on the first line give $R_B'
= 4 C \as(\ptjv)/\pi \ln 1/v$;
the second set of round brackets on that line gives $-\ln R$ (for now
we neglect the $\order{1}$ contribution from the ill-defined upper
limit in $\Delta_{ab}$); the last line gives $\ln (1/\max(z,1-z))$.
We are therefore left with
\begin{multline}
  \label{eq:cF1correl-smallR-simple-later}
  \cF^\mathrm{correl} = 
  R_B' \left(\ln \frac{1}{R} + \order{1}\right) \frac{\as(\ptjv)}{\pi} \int_0^1
   dz \frac1{2!} \left( P_{gg}(z) + 2 n_f P_{qg}(z) \right) \ln \frac{1}{\max(z,1-z)}\,.
\end{multline}
This is straightforward to evaluate and gives
\begin{subequations}
  \begin{align}
    \cF^\mathrm{correl} &= R_B' \frac{\as(\ptjv)}{\pi} \left( C_A \frac{12\pi^2 +
        132 \ln 2 - 131}{72} + n_f T_R \frac{23-24\ln 2}{36}\right)
    \left(\ln
      \frac1R + \order{1}\right)\,,\\
    &\simeq R_B' \frac{\as(\ptjv)}{\pi} \left( 1.09626 C_A + 0.176791 n_f T_R
    \right) \left(\ln \frac1R + \order{1}\right)\,.
  \end{align}
\end{subequations}

\begin{figure}
  \centering
  \includegraphics[height=0.38\tw]{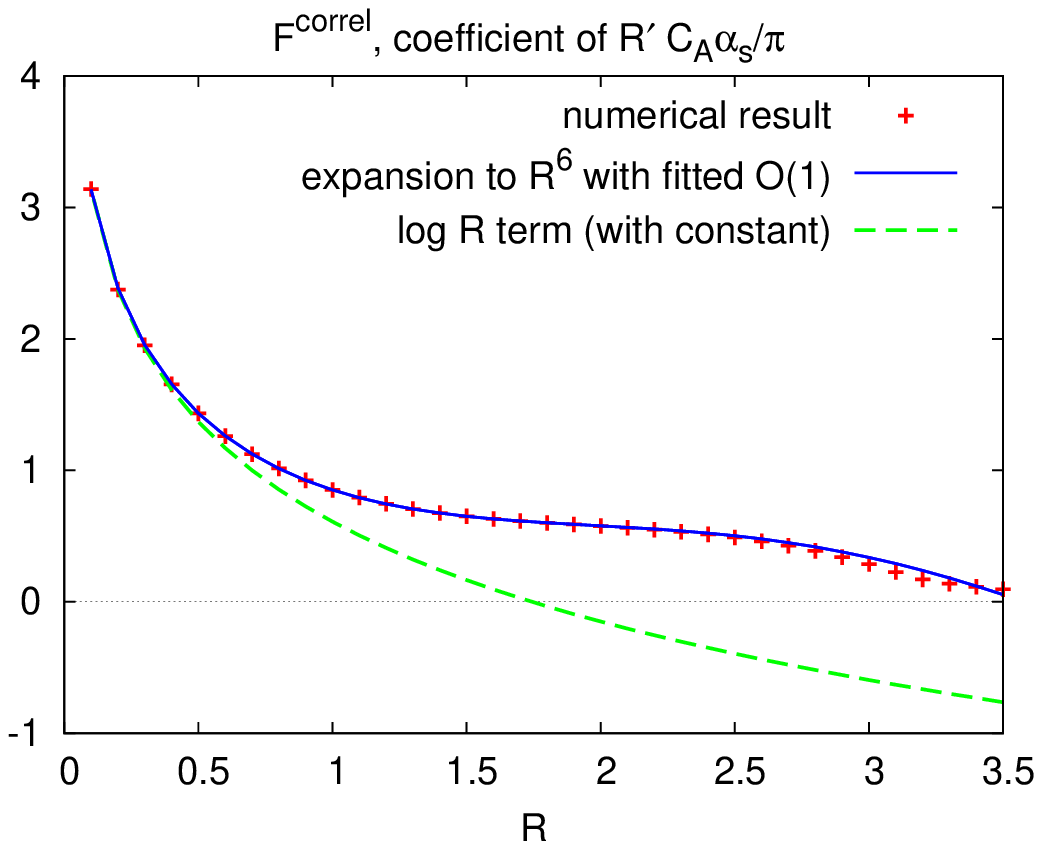}
  \hfill
  \includegraphics[height=0.38\tw]{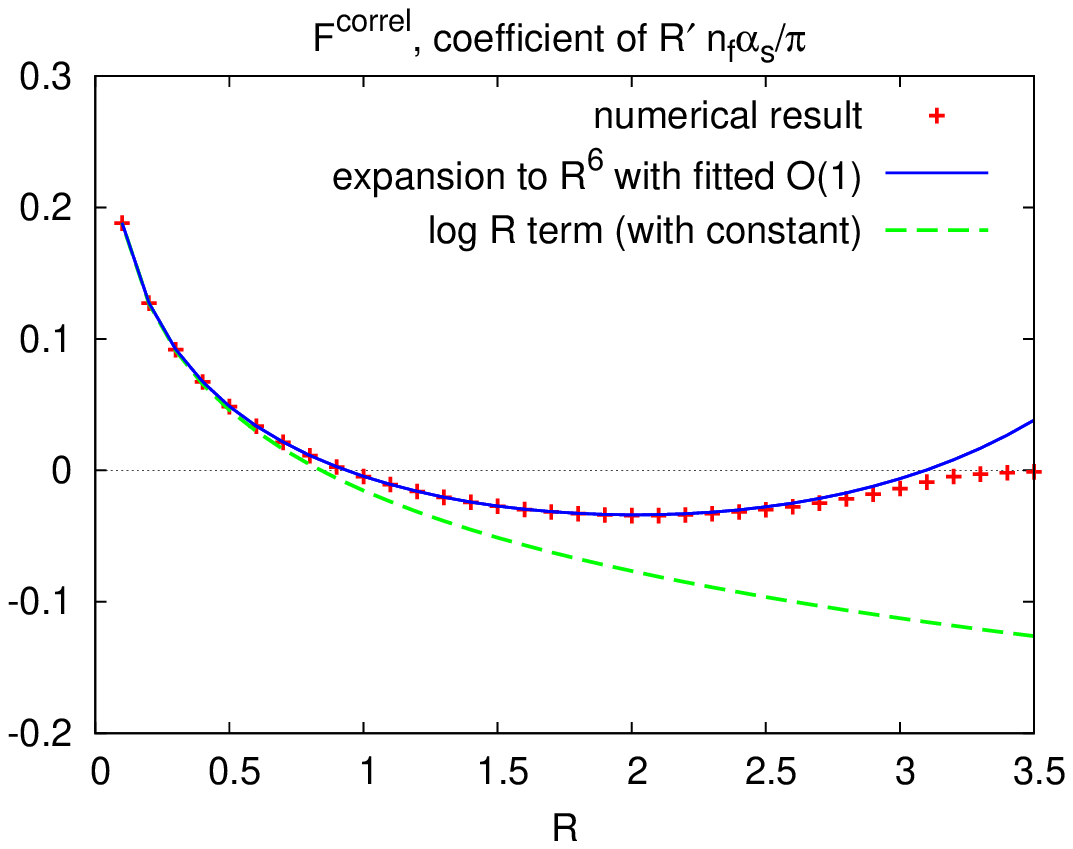}
  \caption{Comparison of the numerical and small-$R$ analytical
    determinations of the $C_A$ and $C_F$ pieces of
    $\cF^{\text{correl}}$.} 
  \label{fig:Fcorrel-CA-nf}
\end{figure}

For a complete evaluation of the $\cF^{\text{correl}}$, we start again
from Eq.~(\ref{eq:cF1correl-just2}). We multiply and divide
${\widetilde \cM}_{gg,rc}^2(k_a,k_b)$ by a factor $\cM_{g,rc}^2(k_a)
\cM_{g,rc}^2(k_b)$, 
and make use of
\begin{equation}
  \label{eq:one-particlephase-M2}
  [dk]\cM_{g,rc}^2(k) = \frac{2\as(k_t) C}{\pi} \frac{ dk_t}{k_t} \frac{d\phi}{2\pi} dy\,.
\end{equation}
We then replace the integration measure $dk_{t,b}/k_{t,b} d\phi_b dy_b$
with  $d\zeta/\zeta d\Delta\phi d\Delta y$
where  $\zeta = k_{t,b}/k_{t,a}$, $\Delta\phi = \phi_{b} - \phi_a$,
$\Delta y = y_b - y_a$.
This gives us
\begin{multline}
  \label{eq:cF1correl-general-simple}
  \cF^\mathrm{correl} = 
   \frac1{2!} \int 
   \left(\frac{2\as C}{\pi}  \frac{d\phi_a}{2\pi} dy_a \right)
   \frac{2\as C}{\pi} \int_0^{\infty}\frac{ d\zeta}{\zeta} \int_{-\pi}^{\pi}\frac{d\Delta\phi}{2\pi}
   \int_{-\infty}^{\infty}d\Delta y \,
   \times \\ \times
   \frac{{\widetilde \cM}_{gg}^2(k_a,k_b)}{\cM_{g}^2(k_a) \cM_{g}^2(k_b)} 
   \frac{ dk_{t,a}}{k_{t,a}} \left[\Theta(v - V(k_a,k_b)) - \Theta(v - V(k_a+k_b)) \right] \,.
\end{multline}
As before, the first factor in round brackets on the first line will
integrate to give $R_B'$.
The difference of $\Theta$ functions on the second line will be
non-zero only when the two partons are separated by $\Delta\phi^2 +
\Delta y^2 > R^2$ (this statement holds only for a recombination
scheme such as the $E$-scheme, which directly sums 4-vectors).
In the ratio of matrix elements we drop the ``$rc$'' subscript, since
at our accuracy running coupling effects are entirely accounted for in
the scale choice that we will make, $\as(\ptjv)$, for the explicit
factors of $\as$ (cf.\ the discussion in the small-$R$ limit).
Observing that $k_{t,ab}^2/k_{t,a}^2$ is independent of $k_{t,a}^2$ and
equal to $(1 + \zeta^2 + 2 \zeta \cos \Delta \phi)$, we can then perform
the $k_{t,a}$ integration together with the $\Theta$-function
constraints in the square brackets to yield a factor
\begin{equation}
    \frac12 \ln \left[\frac{1 + \zeta^2 + 2 \zeta \cos \Delta \phi}{\max\{\zeta^2,1\}}\right]\,.
\end{equation}
Our result for $\cF^{\text{correl}}$ is then
\begin{multline}
  \label{eq:cF1correl-general-final}
  \cF^\mathrm{correl} = 
   R_B' \,   \frac{2\as(\ptjv) C}{\pi} 
   \int_0^{\infty}
   \frac{ d\zeta}{\zeta} \int_{-\pi}^{\pi}\frac{d\Delta\phi}{2\pi}
   \int_{-\infty}^{\infty} d\Delta y \;
   \Theta(\Delta\phi^2 + \Delta y^2 - R^2) \,\times
   \\ \times
   \frac1{2!} \frac{{\widetilde \cM}_{gg}^2(k_a,k_b)}{\cM_{g}^2(k_a) \cM_{g}^2(k_b)} 
   \frac12  \ln \left[\frac{1 + \zeta^2 + 2 \zeta \cos \Delta \phi}{\max\{\zeta^2,1\}}\right]\,.
\end{multline}
In both $R_B'$ and the explicit factor  of $\as$, the coupling is
evaluated at scale $\ptjv$, as before.
The derivative with respect to $R$ of
Eq.~(\ref{eq:cF1correl-general-final}) can be straightforwardly
evaluated as an expansion in powers of $R$.
Integrating that expansion gives us a result for $\cF^\mathrm{correl}$
that is missing a constant of integration.
This constant can easily be determined through a numerical integration
of Eq.~(\ref{eq:cF1correl-general-final}), which also allows for a
check of the range of validity of the power series in $R$.
The result that we obtain is
\begin{multline}
  \label{eq:Fcorrel-final}
  \cF^{\text{correl}} =  R_B' \,   \frac{\as(\ptjv) }{\pi}  \bigg[
  \frac{\left(-131+12 \pi ^2+132 \ln 2\right)
    C_A}{72}  \ln \frac{1.74}{R}
    + \frac{(23-24 \ln 2) n_f}{72}
  \ln \frac{0.84}{R}
  \\
  +\frac{\left(1429+3600 \pi
        ^2+12480 \ln 2\right) C_A+(3071-1680 \ln 2)
      n_f}{172800} R^2
  \\
  +\frac{\left(-9383279-117600 \pi
        ^2+1972320 \ln 2\right) C_A+2 (178080 \ln 2-168401)
      n_f}{406425600} R^4
  \\
  +\frac{(74801417-33384960 \ln 2)
      C_A+(7001023-5322240 \ln 2)
      n_f}{97542144000} R^6
  +O\left(R^8\right)
  \bigg]\,.
\end{multline}
This result is compared to the full numerical determination in Fig.~\ref{fig:Fcorrel-CA-nf}.

\subsection{Independent emission}

In the case of the jet observables that we have been discussing, 
the starting point for the independent-emission correction is
\begin{multline}
  \label{eq:cF1indep}
  \cF^\mathrm{indep} = \exp\left(-\int_{\epsilon v}^v [dk]
    \cM^2_{g,rc}(k)\right) \times\\\times \sum_{n=0} \frac{1}{n!} \left(
    \prod_{i=1}^n
    \int_{\epsilon v} [dk_i]  \cM^2_{g,rc}(k_i) \Theta(v - V(k_i))\right)
   \frac1{2!} \int [dk_a][dk_b] 
   \cM^2_{g,rc}(k_a) \cM^2_{g,rc}(k_b) \times\\
  \times \left[\Theta(v - V(k_a,k_b)) - 
    \Theta(v - V(k_a))\Theta(v - V(k_b)) 
     \right] \,.
\end{multline}
The second term in square brackets on the last line corresponds to the
approximation made in obtaining $\cF = 1$ at NLL; the first term
corresponds to the actual value of the observable.
The evaluation of this formula largely follows the working given above
in the correlated emission case, with the main difference that now the
only region that contributes is that where the two gluons have
$\Delta_{ab} < R$. 
The result for $R<\pi$ is 
\begin{subequations}
  \begin{align}
    \label{eq:cF1indep-result}
    \cF^{\text{indep}} &= - R_B' \frac{\as(\ptjv) C}{\pi} \frac{1}{2!}\int_0^R
    \Delta_{ab} d\Delta_{ab} \int_0^{2\pi} \frac{d\psi}{2\pi} \int_0^\infty
    \frac{d\zeta}{\zeta} \ln\left[\frac{1
    + \zeta^2 + 2\zeta \cos(\Delta_{ab} \cos\psi)}{\max\{\zeta^2,1\}}\right]\\
    & = R_B' \frac{\as(\ptjv) C}{\pi} \left(-\frac{\pi^2 R^2}{12} +
      \frac{R^4}{16} \right)\,.
  \end{align}
\end{subequations}

\end{document}